\documentclass[12pts]{article}
\usepackage{todonotes}
\usepackage{timet}
\usepackage{bm}
\usepackage{color}
\usepackage{amsmath}
\usepackage{graphicx}
\usepackage{subfig}
\usepackage[utf8]{inputenc}
\usepackage{siunitx}
\usepackage{color,soul}
\usepackage{multibib}

\usepackage{tablefootnote} 

\newcites{M}{Methods References}

\usepackage{verbatim}

\usepackage{lineno}

\usepackage[]{hyperref}

\newcommand{\beginsupplement}{%
  \setcounter{table}{0}
  \setcounter{figure}{0}
  \renewcommand{\thetable}{S\arabic{table}}%
  \renewcommand{\thefigure}{S\arabic{figure}}%
  \renewcommand{\theHtable}{Supplement.\thetable}%
  \renewcommand{\theHfigure}{Supplement.\thefigure}
}

\renewcommand{\figurename}{\textbf{Fig.}}
\bibliographystyleM{Science}

\makeatletter
\let\saved@includegraphics\includegraphics
\AtBeginDocument{\let\includegraphics\saved@includegraphics}
\makeatother

\begin{document}
\baselineskip24pt
\title {Fault-size dependent fracture energy explains multi-scale seismicity and cascading earthquakes}

\author {Alice-Agnes Gabriel$^{1,2}$, Dmitry I. Garagash$^{3}$, Kadek H. Palgunadi$^{4}$, P. Martin Mai$^{4}$
  \\
  \normalsize{$^{1}$ Scripps Institution of Oceanography, University of California San Diego, La Jolla, USA}\\
  \normalsize{$^2$ Department of Earth and Environmental Sciences, Ludwig-Maximilians-Universität München, Munich, Germany}\\
    \normalsize{$^{3}$ Department of Civil and Resource Engineering, Dalhousie University, Halifax, Canada} \\
        \normalsize{$^{4}$ Physical Science and Engineering, King Abdullah University of Science and Technology, Thuwal, Saudi Arabia}}

\label{firstpage}
\maketitle
\textit{\large{This is a non-peer reviewed preprint.}}
\section*{Summary}
Earthquakes vary in size over many orders of magnitude, yet the scaling of the earthquake energy budget remains enigmatic.
We propose that fundamentally different ``small-slip'' and ``large-slip'' fracture processes govern earthquakes. 
We combine seismological observations with a physics-based mechanical earthquake model under flash-heating friction. We find that dynamic weakening and restrengthening effects are non-negligible in the energy budget of small earthquakes and establish a simple linear scaling relationship between small-slip fracture energy and fault size.
We use supercomputing to apply this scaling and unveil volumetric ``Mode-4'' earthquake cascades involving  $>700$ multi-scale fractures within a fault damage zone, capable of dynamically triggering large earthquakes. 
Our findings provide an intuitive explanation of seismicity across all scales with important implications for comprehending earthquake nucleation and multi-fault rupture cascades.

\section{Introduction}
\label{introduction}
\begin{figure*}
\includegraphics[width=\textwidth]{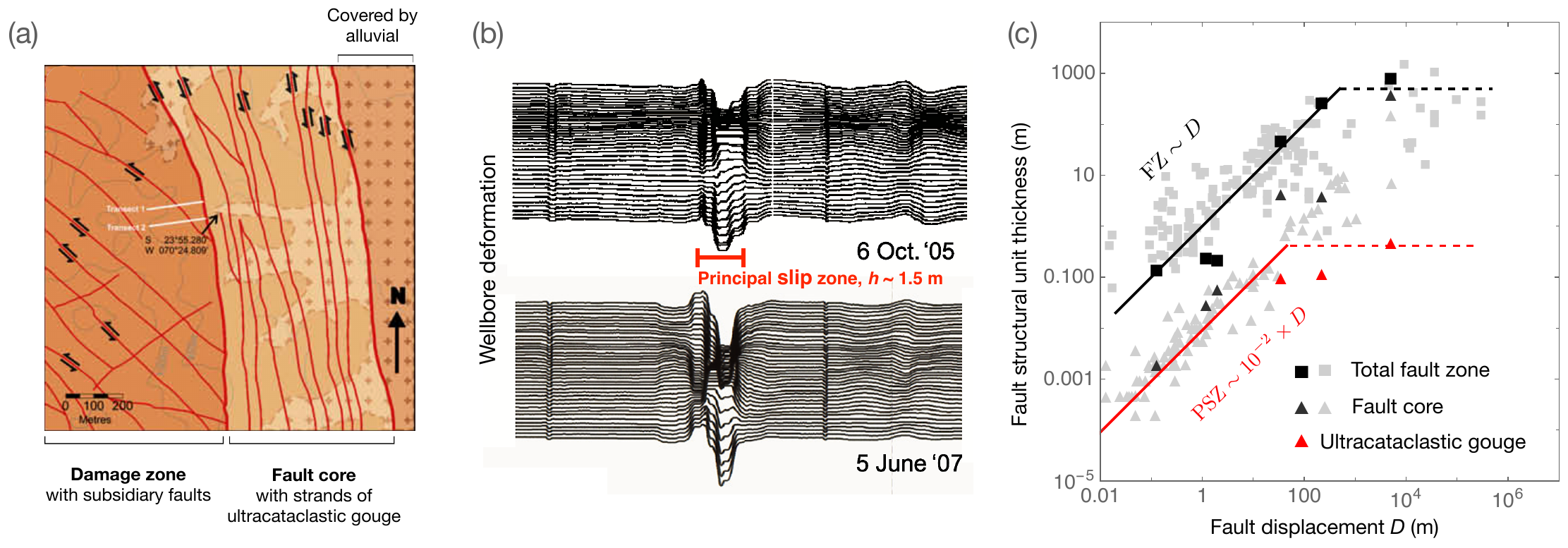}
\caption{(a) Structure of a large fault zone (adapted from \cite{MitchellFaulkner2009}). Geological map of the Caleta Coloso fault, showing a $\approx$400~m thick fault core with $\approx$1~m thick ultracataclastic gouge strands and a $\approx$200~m thick fault damage zone abating the core and hosting subsidiary faults and fractures. 
(b)  The ultracataclastic gouge zone of the San Andreas Fault intersected by the SAFOD wellbore, a principal slip zone with evidence of interseismically delocalized slip. Black lines are the radial deformation of the wellbore casing at various azimuthal directions along its circumference. Deformation is distributed over the entire intersected gouge layer thickness of$\approx$1.5~m (adapted from \cite{Zoback2011}). 
(c) Approximately linear scaling of the thicknesses of the total fault zone ($FZ$) and principal slip zone ($PSZ$) with total fault displacement $D$. 
Both scaling relations saturate when faults exceed $\approx$100~m (PSZ) and $\approx$300~m (FZ) of total fault displacement (dashed lines).
The $PSZ$ is here defined as an ultracataclastic gouge layer within the fault core ($FC$) of large faults or as the entire fault core for smaller, more immature faults. 
We show the thicknesses of $FZ$ (black squares) and $PSZ$ ( $FC$ as black triangles and ultracataclastic gouge as red triangles) for six faults of varying size from \cite{MitchellFaulkner2009}, as well as a larger compilation of $FZ$ (grey squares) from \cite{SavageBrodsky2011} and $FC$ (grey triangles) from \cite{marrett1990kinematic,robertson1987fault}.}
  \label{fig:1}
\end{figure*}

Natural fault zones are multi-scale systems spanning millimeters to thousands of kilometers in fracture and fault lengths (Fig.\ref{fig:1}a, \cite{ChesterEvans93,MitchellFaulkner2009,Cocco2023}). 
Fault cores of large faults are embedded in fault damage zones that include subsidiary faults, enclosed lenses of highly fractured material, and distributed macro-fractures.
Within the fault core, strain is accommodated within one or several principal slip zones of highly-comminuted, ultracataclastic gouge of centimeter-to-meter thickness 
(Fig. \ref{fig:1}b), and may co-seismically localize to a sub-mm `slip surface' \cite{ChesterEvans93,Chester98,DePaola2008,SavageBrodsky2011}.
This structural complexity is reduced for smaller faults where a single fault core is embedded in a thin fault damage zone 
\cite{MitchellFaulkner2009,SavageBrodsky2011}. 
Observations suggest that the thickness of fault zones ($FZ$) and principal slip zones ($PSZ$) both scale approximately linearly with total fault displacement $D$, as $FZ \sim D$ and  $PSZ \sim 10^{-2} D$  (Fig.\ref{fig:1}c). The empirical scaling relation $D \sim 10^{-2} L_f$, suggests an equivalent scaling of $FZ$ and $PSZ$ with fault length $L_f$ \cite{Scholz2019book}.

Despite their structural complexity, natural faults and fractures may generate earthquakes spanning many orders of magnitude ($-4<M<9$, with $M$ being an earthquake magnitude, \cite{Cocco2023,Abercrombie2021,Kwiatek10}).  
Earthquake science suffers from paucity of observations, which are sparse in time and space, thereby limiting our ability to measure small-scale processes and to understand the underlying physics of earthquakes \cite{BenZion2022}.
In addition, the dynamics of how faults slip is a problem that is mostly unsolvable analytically. 
However, simple empirical scaling relations connect small and large faults and the earthquakes occurring on them with models of elliptical cracks and linear-elastic fracture mechanics \cite{udias2014source}. 
Empirical fault constitutive relations, i.e., friction laws \cite{Scholz2019book} informed from small-scale laboratory experiments, are useful to describe co-seismic fault weakening that controls earthquake nucleation, dynamic slip evolution, and rupture arrest. 
Their parameterization differs by up to several orders of magnitude when inferred from laboratory experiments versus observations from real earthquakes \cite{Cocco2023,mikumo2003stress} which may be due to the scale-dependence of fault structural complexity (Fig. \ref{fig:1}) or fault frictional weakening processes at coseismic slip scales \cite{diToro2011}. 

The energy budget of earthquake rupture (Fig.~\ref{fig:2}) describes the dissipation of the released energy during the dynamic evolution of fault stress with slip and the radiation of seismic waves \cite{kostrovDas1988}. Its components are, however, difficult to disentangle and quantify from observations. Specifically, their potential scale-dependence or fault-invariant character remains debated \cite{Cocco2023,Ke2022}.
Fracture energy $G$, a fundamental fault property, balances the release of potential energy $\Delta W$ with several dissipative mechanisms. 
 If the rate at which the elastic energy of the pre-stressed host rock released at the rupture front  matches the fracture energy (e.g., \cite{KanamoriHeaton2000}), faults or fractures of any size may generate earthquakes.
Fracture energy fundamentally affects all aspects of the co-seismic rupture process, including the nucleation of small and large events, radiation of potentially destructive seismic waves, and rupture arrest.
However, only a part of the fracture energy, named $G'$
is possibly inferrable from seismological observations \cite{Abercrombie2021,Lambert2021}.
Other parts, including a multitude of intensely debated co-seismic fault weakening and re-strengthening processes, remain largely undetectable \cite{Madariaga76, Tinti2005}. 
 Specifically, dynamic rupture complexity, including self-healing slip pulses \cite{heaton1990evidence} and arresting cracks \cite{kostrovDas1988} leading to stress ``under- or overshoot'' \cite{Madariaga76}, i.e., fault stresses remaining lower or above the level that would be expected based on the co-seismic slip, are commonly omitted \cite{Abercrombie2021}.
 
 $G'$  has been inferred to increase with earthquake slip \cite{Cocco2023,Abercrombie2021}, and several hypotheses have been proposed to explain this key observation, including 
 (i) the effect of continuous, possibly fault-invariant, co-seismic weakening with slip on all scales (for example, due to thermal pressurization of pore fluids or flash-heating \cite{ViescaGaragash15,brantut2017FH,paglialunga2022FH});
 (ii) fault properties such as fault size, fault maturity, or pre-stress heterogeneity \cite{Scholz2019book,Ke2022} that determine fracture energy scaling.
An important implication inherent to most previous hypotheses is that the seismologically inferred $G'$ may significantly underestimate the total fracture energy $G$ \cite{Abercrombie2021}. 

\section{The earthquake energy budget including dynamic restrengthening}

During earthquake rupture, the total strain energy release per unit of the ruptured surface at a given point of the fault is given by $\Delta W=(\tau_{0}+\tau_{1})\delta/2$ (Fig. \ref{fig:2}a), where $\tau_0$ and $\tau_1$ are the initial and final shear stresses, respectively.
The energy release can be further divided into several components forming the earthquake energy budget \cite{KanamoriHeaton2000}: $\mathcal{E}_s$, the energy radiated by seismic waves; $G$, the fracture energy; $W_r$, the restrengthening work; and $F$, the frictional heat (Fig. \ref{fig:2}a). This energy budget can be averaged over the fault area (e.g., \cite{noda2013stressdrop}) of the entire rupture, which we consider in this study. 

The fracture energy $G$ represents the breakdown work required to reduce fault strength from its peak value, $\tau_p$, at the onset of slip, to a minimum dynamic strength, $\tau_d$, over a critical slip distance $\delta_c$ (Fig. \ref{fig:2}a). 
$G$ is defined as 
$G=\int_{0}^{\delta_c}(\tau(s)-\tau_{d})ds$ and can be expressed from the energy budget as:
\begin{equation}
G=\underset{G'}{\underbrace{\Delta\tau\,\delta/2-\mathcal{E}_{s}}}+\underset{G_{u/o}} {\underbrace{(\Delta\tau_{d}-\Delta\tau)\delta-W_{r}}} \,.
\label{Eq:G}
\end{equation}
with $\delta$ the co-seismic fault slip, $\Delta\tau_d = \tau_p-\tau_d$ being the dynamic stress drop and $\Delta\tau$ the static stress drop. Only a part of $G$, $G'$ can be inferred from seismological observations \cite{Abercrombie2021}, while using kinematic or dynamic finite-source inversion of larger earthquakes allows estimating total $G$ \cite{Tinti2005,Gallovic2019b}, but all with considerable uncertainties. 
We here introduce $G_{u/o}=G-G'$, which comprises  
the effects of dynamically evolving fault stress over- or undershoot, $(\Delta\tau_{d}-\Delta\tau)\delta$ and $W_r=\int_{\delta_{c}}^{\delta}\left(\tau(s)-\tau_{d}\right)ds$.

An accurate estimation of fracture energy $G$ is crucial for understanding earthquake physics.
But the seismologically inferred fracture energy $G'$ approximates the true fracture energy $G$ only under the assumption that the effects of dynamic restrengthening work and fault stress under- or overshoot are negligible which is commonly assumed, specifically for small earthquakes \cite{Abercrombie2021}. 
Numerical solutions for crack-like earthquake sources without dynamic restrengthening \cite{Madariaga76} indeed show only a small dynamic overshoot, which have motivated assuming $G'=G$ to infer the fracture energy of small events, such as aftershock sequences and borehole seismicity, in most previous studies \cite{Abercrombie2021,ViescaGaragash15}. 
Non-negligible $G_{u/o}$ is a common characteristic of physics-based earthquake models and may significantly impact fracture energy estimates but its effects remain largely debated \cite{Lambert2021,Noda09,Rice06,PlattViescaGaragash15}. 

\section{A physics-based correction of the observed fracture energy}
\label{sec:3}
To reconcile seismological observations with earthquake physics, we explicitly quantify $G_{u/o}$ as the contribution of coseismic restrengthening and stress under- or overshoot \cite{Madariaga76} to fracture energy in a physics-based and data-driven manner. Challenging the common assumption that $G\approx G'$ allows us to accurately estimate a scaling of minimum $G$ with ruptured fault size.
We develop a physics-based correction of the observationally inferred $G'$ from a new mechanical model of co-seismic restrengthening. 
We then relate the physics-based maximum dynamic stress drop $\Delta\tau_{d}$ to the observationally inferred static stress drop $\Delta\tau$ to correct $G'$ for $G_{u/o}$.

We derive an analytical description of a circular crack-like dynamic rupture of size $R$ driven by rapid flash-heating frictional weakening at high slip velocity \cite{Noda09,goldsby2011flash}.
We approximate dynamic earthquake rupture as a self-similar singular Kostrov solution \cite{kostrovDas1988}, accounting for time-varying rupture speed and stress drop.
By evaluating the stress distribution during the rupture and tracking the slip history, we derive analytical expressions for $G_{u/o}$ and its components (shown in Fig. \ref{fig:2}b).

We find that the average $G_{u/o}$ can be expressed solely in terms of rupture size $R$, as
\begin{equation}
G_{u/o}=0.4393\times\tau_{*}R\quad\text{with}\quad\tau_{*}=(\tau_{LV}-\tau_{w})\frac{V_{w}}{v_{r}} \,
\end{equation}
where the dynamic stress quantity $\tau_*$ is defined from the flash-heating breakdown strength $(\tau_{LV}-\tau_{w})$, with $\tau_{LV}$ being the low-velocity steady-state dynamic stress and $\tau_{w}$ the fully-weakened flash-heating dynamic stress, the minimum flash-heating slip velocity $V_w$ and rupture front speed $v_r$. 
Similarly, an alternative expression for $G_{u/o}$ in terms of average stress drop $\Delta\tau$ and slip (see Supplementary Materials) reveals its independence from rupture propagation details.

By considering the dynamic contributions of co-seismic restrengthening and stress under- or overshoot effects, we show that in contrast to common assumptions, $G_{u/o}$ does play a significant role in the earthquake energy budget.
Quantitatively comparing $G_{u/o}$ to the average $\frac{1}{2}\Delta\tau\,\delta$, which represents the positive portion of $G'$ (Eq. \ref{Eq:G}), we see that $G_{u/o}$ is negligible for large stress drop events but becomes important and even dominant for events with stress drops comparable or smaller than the re-strengthening scale $\Delta\tau_*\sim\sqrt{\mu\tau_*}$  (Fig. \ref{fig:2}b). We estimate that critical value to be $\sim 5$~MPa for a plausible model parametrization (see Supplementary Materials).

\section{Linearly scale-dependent fracture energy}
\label{sec:4}

Our key results are illustrated in Fig. \ref{fig:3}. 
We utilize our physics-based estimate of $G_{u/o}$ to obtain the full fracture energy $G=G'+G_{u/o}$ from seismologically inferred observations that are now corrected for coseismic rapid dynamic weakening and restrengthening. As we show in Fig. \ref{fig:3}a, there is an evident steady increase in the corrected fracture energy with ruptured fault size, consistent with the scaling of $G_{u/o}$ that we established, but not with fault slip (Fig. \ref{fig:3}b).
This contradicts previous inferences of fracture energy $G$ rising with both event slip and event size across all scales, such as in the uncorrected inferences of $G'$ (in grey in Fig. \ref{fig:3}).

We here propose that the total fracture energy $G$ comprises two independent and equally important components: 
(i) $G_{c}(R)$, a `small-slip' minimum fracture energy that is an irreducible fault property linked to the ruptured fault size $R$ and which dominates faulting at small-slip scales, 
and (ii) $\Delta G(\delta)$, a `large-slip' and possibly fault-invariant part of fracture energy that increases continuously with co-seismic slip $\delta$, as
\begin{equation}
G=G_{c}(R)+\Delta G(\delta) \,.
\label{eq:GcDG}
\end{equation}

The minimum fracture energy $G_c(R)$ emerges in Fig. \ref{fig:3}a as a linear function of ruptured fault size, with
\begin{equation} \label{eq:GcR400}
G_{c}(R)\approx 400\text{ [Pa]}\times R\,. 
\end{equation}
Events with fracture energy above this lower bound 
may have experienced additional weakening or represent partial ruptures (see Supplementary Materials). At the same time, we observe a clear break in the scaling of $G$ with co-seismic slip in Fig. \ref{fig:3}b. The corrected fracture energy exceeds the prediction of a continued slip-weakening model $G>\Delta G(\delta)$ for small events but agrees at large enough slip. Both findings can be reconciled by earthquakes governed by different weakening mechanisms with distinctive scaling behavior activated at different levels of co-seismic slip, leading to our proposed fracture energy decomposition in Eq. \ref{eq:GcDG}.

Our subsequent analysis assumes that the observed ruptured fault size $R$ is representative of the full fault size, i.e., that a representative subset of the data corresponds to events that have ruptured nearly the entire area of their corresponding faults. We imply that $G_c(R)$ is a local fault property that depends on fault size, which can be explained by a well-localized near-front process zone (Supplementary Materials).
In practice, it is difficult to distinguish full and partial ruptures.
In our model, partial ruptures may manifest as the data points plotted above the $G_c(R)$ line in Fig. \ref{fig:3}a.

\section{Mode-4: Volumetric multi-fault earthquake cascades in 3D dynamic rupture simulations with fault-size dependent fracture energy}
\label{sec:5}

Including a simple linear scaling of the minimum fracture energy with fault size $G_c(R)$ allows us to simulate volumetric earthquakes, occurring as a cascade of ruptures over a multi-scale fracture network and interacting with an embedded main fault (Fig. \ref{fig:4}).
Our models introduce a degree of realism beyond typical earthquake physics models, which often idealize fault zones as infinitesimally thin interfaces with distinct on- versus off-fault rheologies.
These multi-fracture cascades are capable of generating significant ``Mode-4'' earthquakes, consisting of sustained and distributed cascading fracture slip that may dynamically trigger main-fault earthquakes.

Dynamic rupture modeling may involve varying characteristic slip distances across different fault or stress-heterogeneity scales \cite{IdeAochi2005,Cocco2009} to capture rupture growth. 
In our simulations, earthquake rupture dynamics and the potential to dynamically branch or ``jump'' multiple segments are largely controlled by scale-dependent $G_c(R)$ and by 
the dynamic stress drop relative to the maximum dynamic strength reduction, the relative prestress ratio $\mathcal{R}$. 
We introduce a fracture-scale-dependent state evolution slip distance \cite{Noda09}, $L$, directly informed from the scale-dependence of the minimum fracture energy $G_c(R)$ (Eq. \ref{Methods:eq:L_R}, Supplementary Material). We adopt a novel approach by explicitly modeling 3D dynamic earthquake rupture and interaction across 721 multi-scale, partially intersecting fractures and an embedded planar strike-slip which can host ruptures spanning 4 orders of seismic moment magnitudes ($M_w=2\dotsc6$, Fig. \ref{fig:S1}). 
Leveraging High-Performance Computing to incorporate a scale-dependent $L$ in a multi-scale fault zone model enables us to gain insights into the complex mechanisms driving cascading earthquake nucleation, propagation, and arrest (e.g., \cite{Taufiqurrahman2023}). For a detailed analysis using a similar but more realistic listric thrust faulting setup, refer to \cite{palgunadi2022dynamic}.

Fig. \ref{fig:4} shows three cases varying in their prestress conditions and orientation of the maximum ambient horizontal stress ($S_{Hmax}$) relative to the fractures and the main fault, which results in distinctly different rupture dynamics, and hence variations in kinematics and overall earthquake slip patterns but comparable moment magnitudes.
In the first case, fractures are predominantly unfavorably prestressed, leading to dynamic rupture primarily breaking the main fault and inducing off-fault fracture slip, which resembles asymmetric deformation patterns in nature and theoretical models (e.g., \cite{MitchellFaulkner2009,andrews2005rupture,perrin2016,okubo2019dynamics}, Fig. \ref{fig:4}a).
In contrast, the second case, with both the fracture network and the main fault more optimally oriented, facilitates a sustained rupture cascade, branching and jumping across 482 discrete fractures without triggering sustained slip on the main fault (Fig. \ref{fig:4}b). The cascade takes the form of a volumetric rupture pulse \cite{heaton1990evidence}, with a band of fractures actively slipping at any point in time.
We term such events ``volumetric earthquake cascades'': multi-fault ruptures sustained entirely within a 3D fracture network, generating a sizeable Mode-4 event, here, without activating a main fault.
In the third case, with closer-to-critical prestress conditions on the fractures and the main fault, the volumetric earthquake cascade dynamically triggers the main fault after 2.1 seconds, resulting in a compound event with delayed main fault rupture (Movie S3, Fig. \ref{fig:4}c).
A fourth case (Fig. \ref{fig:S2}) illustrates the importance of assuming fault-size-dependent fracture energy. Assuming constant large $G_c$ leads to main fault slip but fails to activate smaller volumetric multi-fracture ruptures.

Despite their complexity, the macroscopic kinematics of Mode-4 rupture cascades (Fig. \ref{fig:S3}, Supplementary Material) highlight distinct, potentially observable features: slow apparent cascading rupture speed  ($\sim 0.65 c_s$, with $c_s$ being the S-wave speed, Fig. \ref{fig:S3}b), despite localized occurrences of supershear rupture speeds; short rise times (Fig. S3b,c);
multi-peak moment rate release corresponding to multiple sub-events of volumetric fracture network slip (Fig. \ref{fig:S3}d); realistic high-frequency seismic radiation and realistic average stress drops, which are elevated for Mode-4 ruptures (Fig. \ref{fig:S3}e--h).

\section{Discussion and Conclusions}

The implications of our fault-size-dependent fracture energy model extend beyond theoretical analyses. It provides a data-consistent and physics-grounded explanation for why the dynamics of earthquakes often involve the activation of interconnected multi-fault systems spanning a variety of spatial scales (e.g., the 1992 Landers, California; 2016 Kaik\={o}ura, New Zealand; and 2023 Kahramanmaraş, Turkey earthquakes). The 2019 Ridgecrest earthquake sequence is a case in point, rupturing a conjugate multi-scale fault system \cite{ross2019}, comprising northeastern and northwestern trending high-angle strike-slip faults.  Our correlation supports the mechanical viability of composite earthquake ruptures occurring as cascades over networks of faults of diverse sizes.

The diverse slip directions observed in off-fault fractures (Fig. \ref{fig:5}a) cannot be fully explained by background or static stress changes alone \cite{xu2020coseismic}, suggesting the presence of dynamic fracture network cascading effects as demonstrated in our models (Fig. \ref{fig:5}b,c). 
High-resolution optical satellite image correlation has confirmed the importance of distributed faulting and diffuse deformation, accounting for up to 50\% of coseismic surface displacement \cite{KlingerSolaineRidgecrest}. This aligns with the off-fault/main-fault slip partitioning in our  model of a volumetric rupture cascade dynamically triggering a main fault (Fig. \ref{fig:4}c.)

Our hypothesis may also shed light on the intriguing observation that far-field focal mechanisms of large earthquakes can be misaligned with their main fault plane \cite{ScottKanamori1985consistency}, as well as with the focal mechanisms of ``volumetric'' aftershocks (e.g., \cite{Hauksson02}). Such misalignments may be attributed to cascading fault zone ruptures driven by fault-size-dependent fracture energy.
Even on smaller scales, such as during the 2016–2019 Cahuilla earthquake swarm \cite{Cochran2023}, volumetric seismicity is complex, potentially indicating multi-fracture network rupture phenomena.

Previous explanations for the scaling of observed fracture energy fail to account for small earthquakes because faults smaller than the nucleation size $R_c$ are theoretically unable to nucleate dynamic rupture.
For example, the continuous co-seismic weakening model under thermal pressurization (red line in Fig. \ref{fig:3}b, \cite{ViescaGaragash15}) falls short in accounting for seismogenesis on faults smaller than $R_c \sim 35$ m (see Supplementary Materials).
In contrast, our proposed scaling provides an intuitive explanation for seismicity observed across all scales, including potential small volumetric seismicity occurring entirely off main faults, such as within subsidiary fracture networks within the damage zone.

Our findings also have implications for the earthquake cycle and the mechanics of natural fault systems.   
We suggest that the redistribution of stresses by cascading ruptures in off-fault fracture networks may assist in the nucleation of larger events and account for dynamic variations in fault strength and stress, even if the main fault is weak, such as in subduction zones \cite{Shreedharan2023}.
 As illustrated in Fig. \ref{fig:5}d, a Mode-4 earthquake initiation model combines aspects of the end-member ``cascade'' and ``pre-slip'' models \cite{mcLaskey2019earthquake}: cascading compound off-fault seismicity, governed by small-slip, scale-dependent $G_c$, leads to the nucleation of a large earthquake, with or without sustained rupture of the main fault. 

The physical mechanisms behind our proposed simple linear scaling of the ``small-slip'' fracture energy with fault size present an intriguing topic for further investigation.
Localization of brittle deformation before and during an earthquake is highlighted in recent laboratory experiments, theoretical models, and statistical analyses of seismicity \cite{Rice06,lockner1991quasi,BenzionZaliapin2020,Platt14,Proctor14,pranger2022RS}. In the context of localization, the minimum fracture energy $G_c$ may be interpreted as the fracture energy of a coseismic localization process. Then, we expect that $G_c$ scales with the thickness of the fault's principal slip zone (Fig. \ref{fig:1}), which itself scales with total fault displacement and fault length (Sec.\ref{introduction}).
Thereby, localization offers a physical mechanism explaining the onset of flash-heating weakening related to the dramatic co-localization drop of fault strength informing $G_c$. Subsequent post-localization slip would favor more efficient pore fluid thermal pressurization \cite{ViescaGaragash15,Rice06} leading to continuing weakening with slip (Fig. \ref{fig:3}b).

In summary, our fault-size-dependent fracture energy model provides an intuitive and comprehensive framework for understanding earthquake complexity across scales. Considering that fundamentally different ``small-slip'' and ``large-slip'' fracture energy components govern earthquakes allows new insights into the mechanisms driving earthquake nucleation, propagation, and natural fault zone interaction and coseismic interconnectivity with fundamental implications for earthquake physics and seismic hazard assessment.

\clearpage
\bibliography{mode_4_refs}

\clearpage 
Harsha Bhat, Samson Marty, Francois Passelegue, and Robert Viesca are thanked for helpful discussions.
We thank Thomas Ulrich and SeisSol's core developers (see \url{www.seissol.org}).
We thank the developers of the following open-source software packages that we used for data analysis and visualization: Obspy \cite{beyreuther2010obspy}, Paraview \cite{ahrens2005paraview} and Matplotlib \cite{hunter2007matplotlib}. 

\subsection*{Funding:} 

Funding for this work was provided by KAUST research grants (BAS/1/1339-01-01 and URF/1/3389-01-01 to P.M.M.), the European Union’s Horizon 2020 Research and Innovation Programme (grant No. 852992 to A.A.G.), Horizon Europe (grant No. 101093038, 101058129, and 101058518 to A.A.G.), the National Aeronautics and Space Administration (grant No. 80NSSC20K0495 to A.A.G.), the National Science Foundation (grant No. EAR-2121666 to A.A.G.), the Natural Sciences and Engineering Research Council of Canada (Discovery Grant 05743 to D.I.G.) and the Southern California Earthquake Center (SCEC awards 22135, 23121 to A.A.G.)

\subsection*{Authors contributions:}
A.A.G. Conceptualization, Data curation, Funding acquisition, Formal Analysis, Methodology, Project administration, Resources, Software, Supervision, Validation, Visualization, Writing – original draft. 
D.I.G. Conceptualization, Data curation, Funding acquisition, Formal Analysis, Methodology, Supervision, Validation, Visualization, Writing – review \& editing. 
K.H.P. Formal Analysis, Methodology, Investigation, Visualization, Writing – review \& editing. 
P.M.M. Funding acquisition, Project administration, Supervision, Writing – review \& editing. 

\subsection*{Competing interests:} 

The authors declare no competing interests.

\subsection*{Data and materials availability:}

\begin{figure*}
\includegraphics[width=\textwidth]{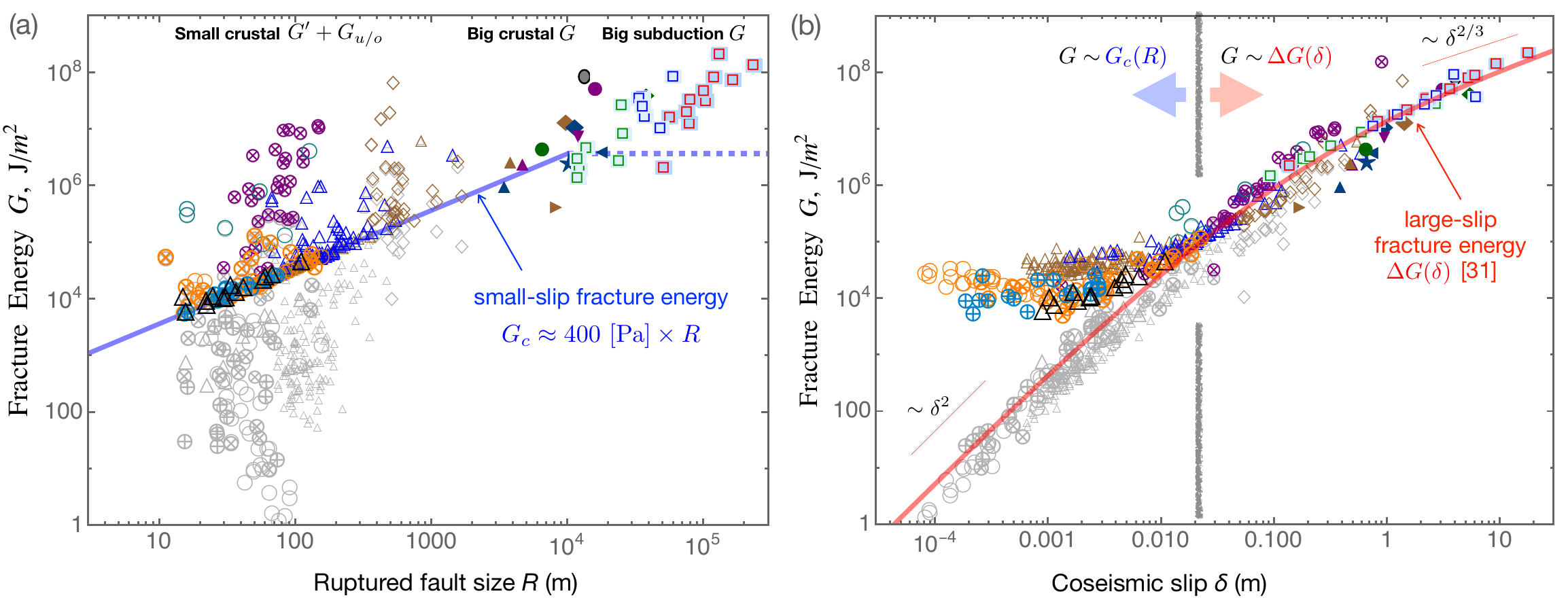}
\caption{(a) Fracture energy $G$ vs. ruptured fault size $R$ for small and large crustal events and subduction earthquakes (colored symbols). Small crustal earthquakes (colored open symbols) include our correction of the previously inferred $G'$ (grey open symbols, \cite{ViescaGaragash15} and references therein) for coseismic rapid dynamic weakening and restrengthening due to flash-heating friction ($G_{u/o}$). For larger crustal and subduction earthquakes, fracture energy $G$ can be estimated and is shown as filled symbols after \cite{Tinti2005, ViescaGaragash15}. These estimates of $G$ are considerably larger than the observational inferences of $G'$ \cite{ViescaGaragash15}. The blue line is the minimum fracture energy $G_{c}(R)\approx 400\text{ [Pa]}\times R$, which we interpret as a small-slip fracture energy. 
(b) Same as (a) but $G$ plotted vs. co-seismic slip $\delta$. Here, the solid red line shows the theoretically predicted fracture energy increase due to thermal pressurization with coseismic slip (see \cite{ViescaGaragash15} and Supplementary Materials).}
\label{fig:3}
\end{figure*}

The inferred fracture energy data presented in Fig. \ref{fig:3} and the corresponding references to the source parameters are available from Supplementary Information Tables S.3 and S.4 of \cite{ViescaGaragash15} for large subduction and crustal events, respectively, and from Table S.5 for small crustal events.
We use the commercial software FracMan Version 7.8 to generate the fracture network.
The version of SeisSol used for the dynamic rupture models in Section \ref{sec:5} is described in \url{https://seissol.readthedocs.io/en/latest/fault-tagging.html#using-more-than-189-dynamic-rupture-tags} with commit version \verb|917250fd| and from the branch SeisSol64FractureNetwork (\url{https://github.com/palgunadi1993/SeisSol/tree/SeisSol64FractureNetwork}). We use a patched version of the open-source meshing software PUMGen which can be cloned from the GitHub branch PUMGenFaceIdentification64bit (\url{https://github.com/palgunadi1993/PUMGen/tree/PUMGenFaceIdentification64bit}). Instructions for downloading, installing, and running SeisSol are available in SeisSol's online documentation at \url{https://seissol.readthedocs.io/}. Instructions for compiling SeisSol are at \url{https://seissol.readthedocs.io/en/latest/compiling-seissol.html}. Instructions for setting up and running simulations are at \url{https://seissol.readthedocs.io/en/latest/configuration.html}. All input and mesh files are available in the Zenodo repository at \url{ https://doi.org/10.5281/zenodo.8108710}.
\clearpage

\begin{figure*}
\includegraphics[width=\textwidth]{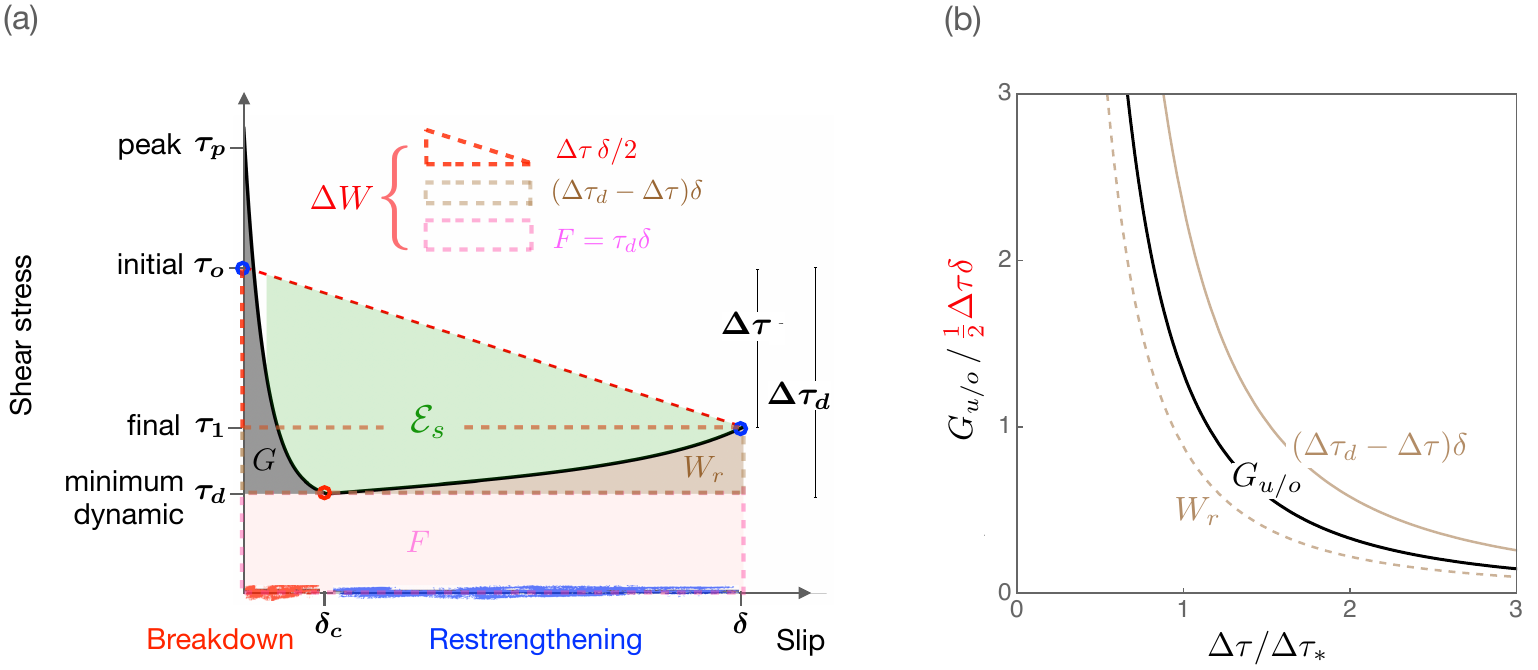}
\caption{ 
(a) Diagram of the total earthquake energy budget, illustrating the balance of the (total) strain energy release $\Delta W$  and several dissipative mechanisms while accounting for dynamic restrengthening: 
$\mathcal{E}_S$ is the energy radiated by seismic waves;
$W_r$ is the restrengthening work;
$F$ is the frictional heat; 
and $G$ is the fracture energy often referred to as the breakdown work. We here split $G$ into the seismologically-inferable $G’=\frac{1}{2}\Delta\tau\delta-\mathcal{E}_s$ and an often omitted component $G_{u/o}=G-G'=(\Delta\tau_d-\Delta\tau)\delta-W_r$. $G_{u/o}$ is associated with the dynamic evolution of fault stress over-/undershoot and restrengthening work during coseismic slip.
 Although the exact partitioning between frictional heat ($F=\tau_{d}\delta$) and total mechanical work (breakdown and strengthening, $G+W_r$) is still a matter of debate \cite{Cocco2023}, we here align with the common assumption that the partitioning is marked by the minimum dynamic fault strength $\tau_{d}$.
(b) Normalized $G_{u/o}=G-G'$, the difference between the fracture energy $G$ and its seismologically observable part $G'$ as a function of the normalized stress drop $\Delta\tau$. These relationships emerge from a newly developed physics-based analytical model of a circular crack driven by flash-heating frictional weakening at the dynamic rupture front and re-strengthening in its wake (see Supplementary Materials).
We illustrate that $G_{u/o}$ constitutes a significant portion of $G$ when the stress drop is smaller or similar to the restrengthening scale $\Delta\tau_*\sim$5~MPa. The beige curves depict the components of  $G_{u/o}=(\Delta\tau_d-\Delta\tau)\delta-W_r$, corresponding to the dynamic stress over-/undershoot and restrengthening, respectively.
}
\label{fig:2}
\end{figure*}

\begin{figure*}
\centering
\includegraphics[width=0.75\textwidth]{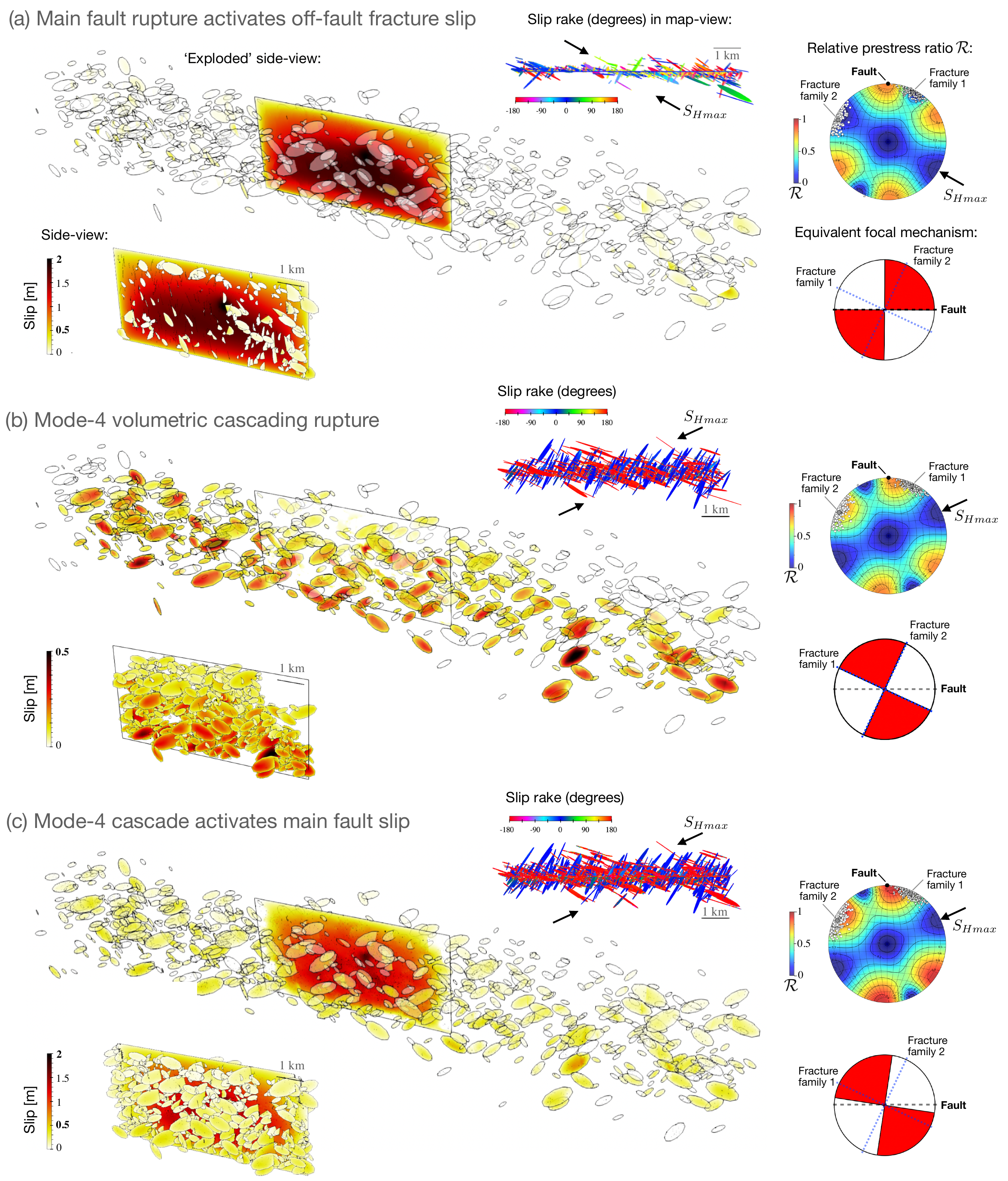}
\caption{
3D dynamic rupture simulations in a multi-scale fracture network with linearly size-dependent fracture energy demonstrate ``volumetric earthquake cascades'' in a fault damage zone interacting with an embedded planar right-lateral strike-slip main fault. Two fracture families form an average angle of $100^\circ$ to each other and angles of $25^\circ$ and $-65^\circ$ with respect to the main fault's strike (Fig. \ref{fig:S1}).
We present three simulations with varying ambient prestress conditions, all initiating dynamic rupture at the same hypocenter. Slip is shown on the fractures and main fault in an 'exploded' side-view, where inter-fracture horizontal distances are scaled by a factor of 3.5, and in an unscaled side-view. The final rake angle of all ruptured portions of the fracture network or main fault is  displayed.
Stereoplots show the fault or fracture-local relative pre-stress ratio ($\mathcal{R}$, which is the ratio of the maximum possible stress drop and frictional strength drop (Supplementary Material) in a lower hemisphere projection. This illustrates optimal and non-optimal orientations with respect to the ambient prestress ($\mathcal{R}_0= 0.8 - 0.95$, the maximum possible value of $\mathcal{R}$) for different maximum compressive ambient stress, $S_{Hmax}$, values. The equivalent point-source focal mechanism illustrates the apparent far-field source mechanisms vs. the orientation of both fracture families and the main fault.
(a) Case 1: Main fault dynamic rupture dynamically activates off-fault fracture network slip (total $M_w=5.96$). Fractures are unfavorably prestressed ($\mathcal{R} < 0.3$), yet 328 out of 721 fractures slip predominantly at the dilatational sides of the main fault (Movie S1). 
(b) Case 2: Mode-4 volumetric cascading rupture sustained within the fracture network (total $M_w=5.64$). Fractures and the main fault are more optimally prestressed ($0.6 < \mathcal{R} \leq 0.8$) due to a different orientation of $S_{Hmax}$, while all other model parameters are the same as in (a). While 482 fractures comprise the rupture cascade, the main fault does not host sustained slip (Movie S2). Slip across the fracture network is determined by intricate interactions of zigzagging rupture fronts and variations in static and dynamic stresses. Coulomb-stress changes due to the evolving slip in the fracture network and dynamic shear and normal stresses transported by seismic waves are jointly driving volumetric rupture cascades. The far-field source mechanism is strike-slip but misaligned with the main fault orientation. The band of actively slipping fractures has about the same ``pulse-width'' as the largest fracture in the network ($\approx$500~m).  The cascade rise time, the duration of slip at a given hypocentral distance within the fault zone, is short compared to the overall rupture duration of the cascade ($\sim10\%$) at all azimuths.
(c) Case 3: Mode-4 cascade dynamically activates main fault slip after 2.1~s (total $M_w=6.00$, Movie S3). Fractures and the main fault are now critically prestressed ($0.8 < \mathcal{R} \leq 0.95$), due to a larger magnitude of $\mathcal{R}$, while all other model parameters are the same as in (b). The volumetric rupture cascade in the fault damage zone contributes 42\% of the total seismic moment release.
}
  \label{fig:4}
\end{figure*}

\begin{figure*}
\includegraphics[width=\textwidth]{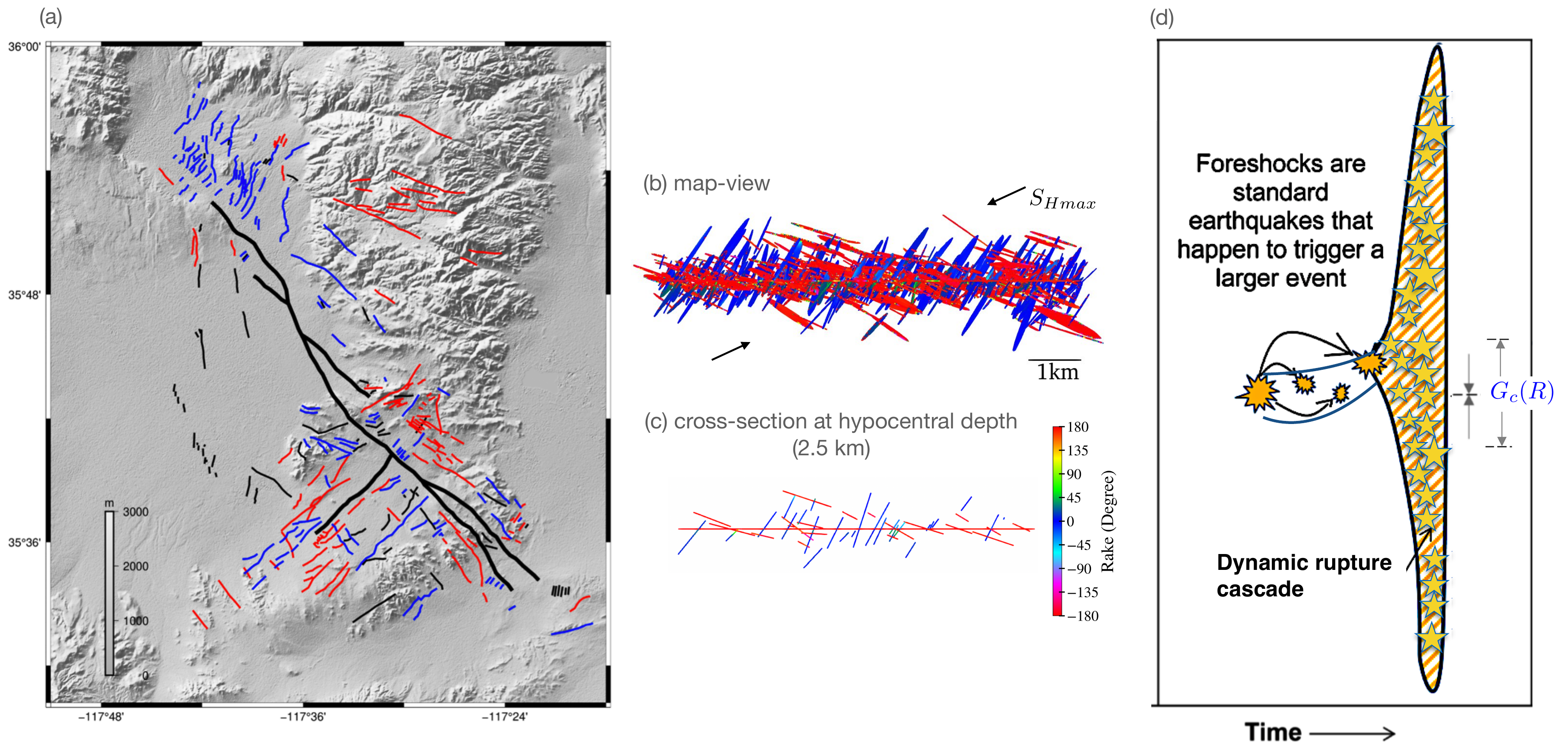}
\caption{
(a)  
Observationally determined map of activated off-fault fractures based on InSAR data for the 2019 Ridgecrest earthquake sequence (adapted from \cite{xu2020coseismic}). Black lines denote the main rupture trace as well as surface fractures that are not predominantly strike-slip. Red lines mark right-lateral, and blue lines mark left-lateral strike-slip motion.
(b) 
Map view showing the variable rake across the complete fracture network in our 3D dynamic rupture model after the volumetric rupture cascade dynamically triggers a main fault (Fig. \ref{fig:4}c).
(c) A horizontal cross-section of (b) at a depth of 2.5~km to align with surface observations in (a).
(d) Illustration of a here discussed Mode-4 earthquake initiation model, modified from \cite{mcLaskey2019earthquake}. Fault-size-dependent minimum fracture energy $G_c$ may drive an off-fault volumetric dynamic rupture cascade within a fracture network. This, in turn, may activate a large main-fault earthquake.}
  \label{fig:5}
\end{figure*}
\clearpage
\beginsupplement
\beginsupplement
\section*{Supplementary Materials}

This file contains legends for Supplementary Movies S1--S3, Supplementary Figures S1-S5, Supplementary Tables S1-S2, Materials and Methods, and References (53-96).

\subsection*{Supplementary Movies}

Movie S1: Evolution of absolute slip rate (m/s) for the dynamic rupture simulation case 1 (Fig. \ref{fig:4}a). \\
Movie S2: Evolution of absolute slip rate (m/s) for the dynamic rupture simulation case 2 (Fig. \ref{fig:4}b). \\
Movie S3 Evolution of absolute slip rate (m/s) for the dynamic rupture simulation case 3 (Fig. \ref{fig:4}c).  \\
Movie S4: Same as Movie S1 in exploded view. \\
Movie S5: Same as Movie S2 in exploded view. \\
Movie S6: Same as Movie S3 in exploded view. \\

\subsection*{Supplementary Figures}

\begin{figure*}
\includegraphics[width=\textwidth]{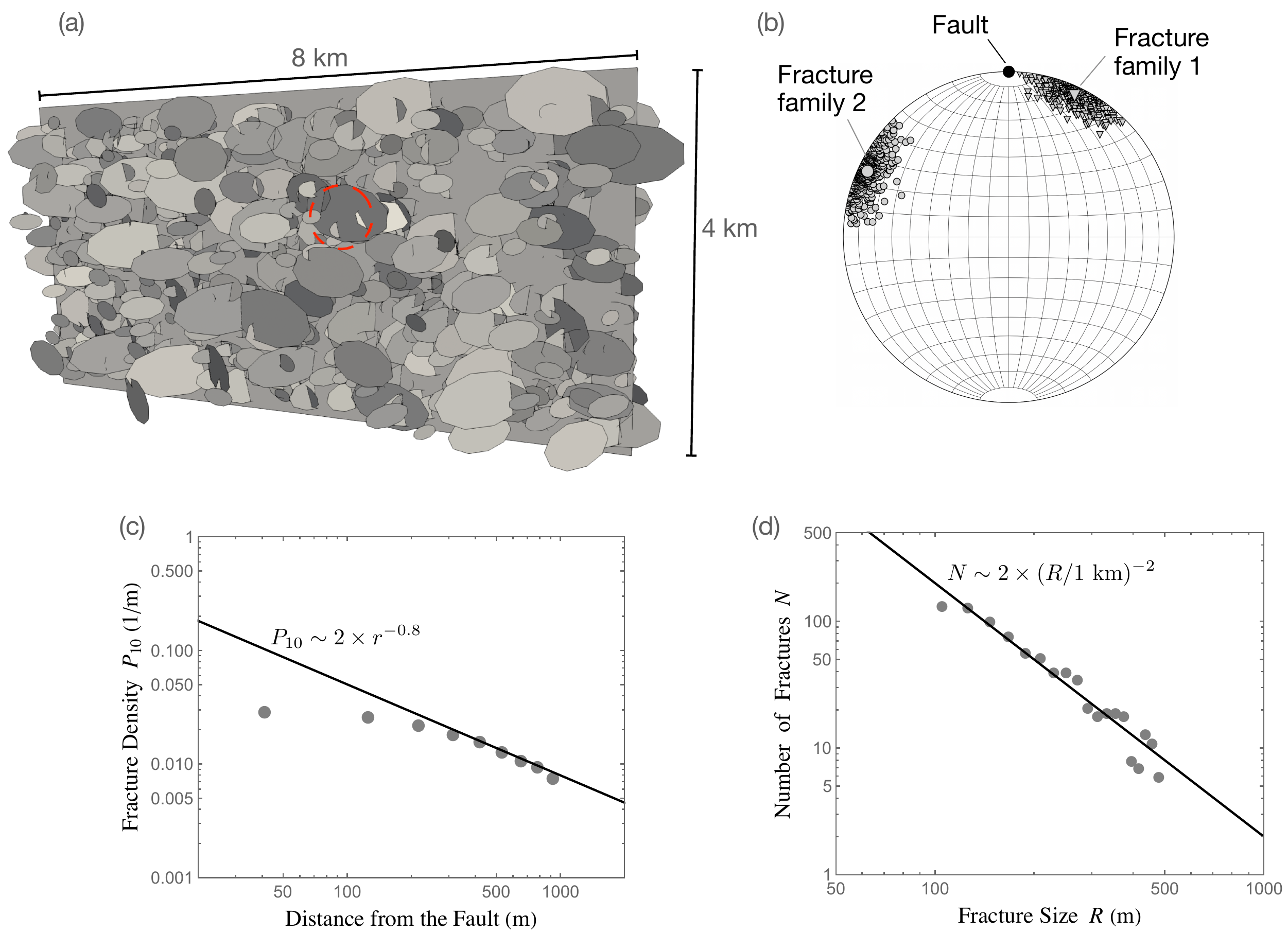}
\caption{ 
(a) Side-view of the 3D dynamic rupture model of a strike-slip fault embedded in a damage zone consisting of a network of 721 multi-scale fractures. Cascading dynamic rupture sequences are initiated by smooth over-stressing within the volume outlined by the red dashed line centered at the hypocenter. The hypocenter coordinates are $x=700$~m, $y=0$~m, and $z=-2500$~m. 
The coordinate system origin with $x=0$~m, $y=0$~m, and $z=0$~m is located at the free surface above the center of the buried vertical fault. The upper edge of the main fault is at  $z=-1000$~m.
(b) Polar view showing the two families of fractures and the main fault in the lower hemisphere, in the direction normal to the fractures and main fault. Grey triangles represent fracture family 1 with an average strike of $25^\circ$ relative to the main fault; the large triangle denotes the mean strike and dip of fracture family 1. Grey dots represent fracture family 2 with a strike of $-65^\circ$; the large dot shows the mean strike and dip of fracture family 2. The black dot at the top indicates the main fault, striking East-West with a dip of $90^\circ$.
(c) Fracture density $P_{10}$ as a function of distance $r$ from the main fault, exhibiting a power law decay $\sim 1/r$ (black line).
(d) Number of fractures versus the fracture size $R$, showing a power law decay $\sim 1/R^2$.
}
  \label{fig:S1}
\end{figure*}

\begin{figure*}
\includegraphics[width=\textwidth]{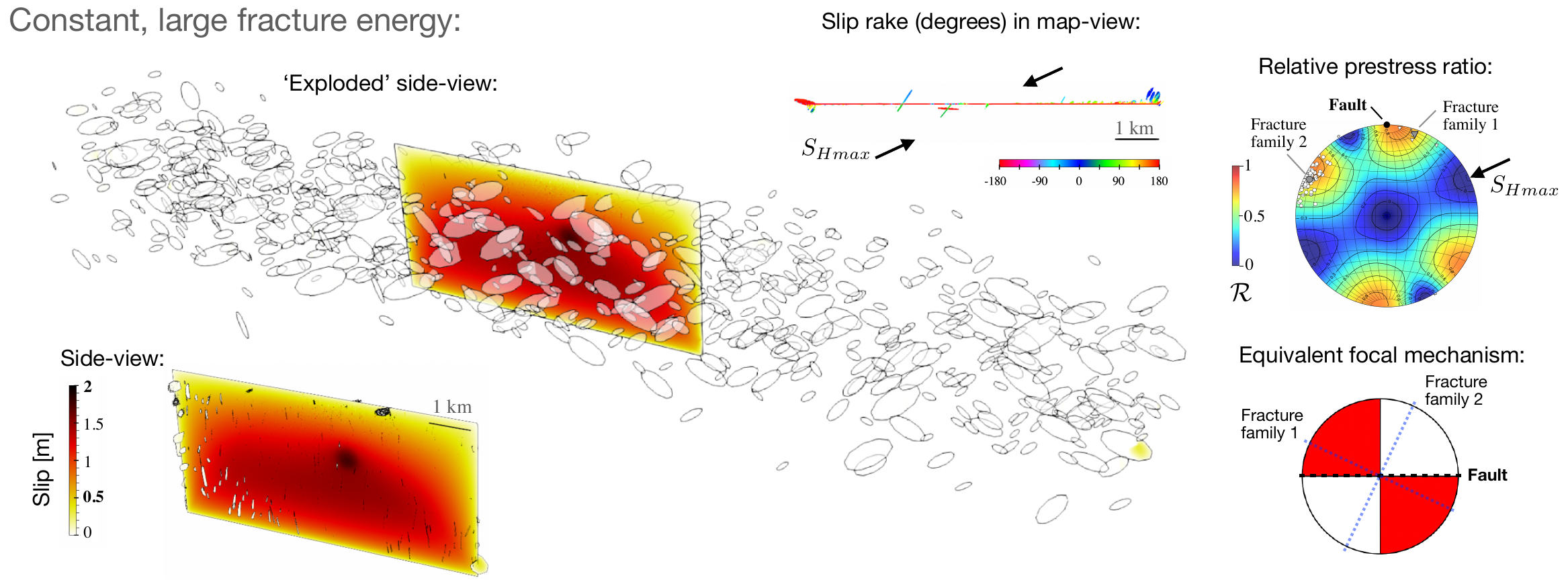}
\caption{Case 4: 3D dynamic rupture simulation in a multi-scale fracture network using a uniformly large value of the fracture energy. The Mode-4 volumetric dynamic rupture cascade is switched off in favor of the main fault rupture. The simulation's parametrization is identical to Case 2, the Mode-4 volumetric cascading rupture in Fig. \ref{fig:4}b, except the here assumed constant value of the state evolution distance $L=0.08$ m (corresponding to the main fault size) and fracture energy $G_c$ across all fractures and the main fault. 
}
  \label{fig:S2}
\end{figure*}

\begin{figure*}
\includegraphics[width=\textwidth]{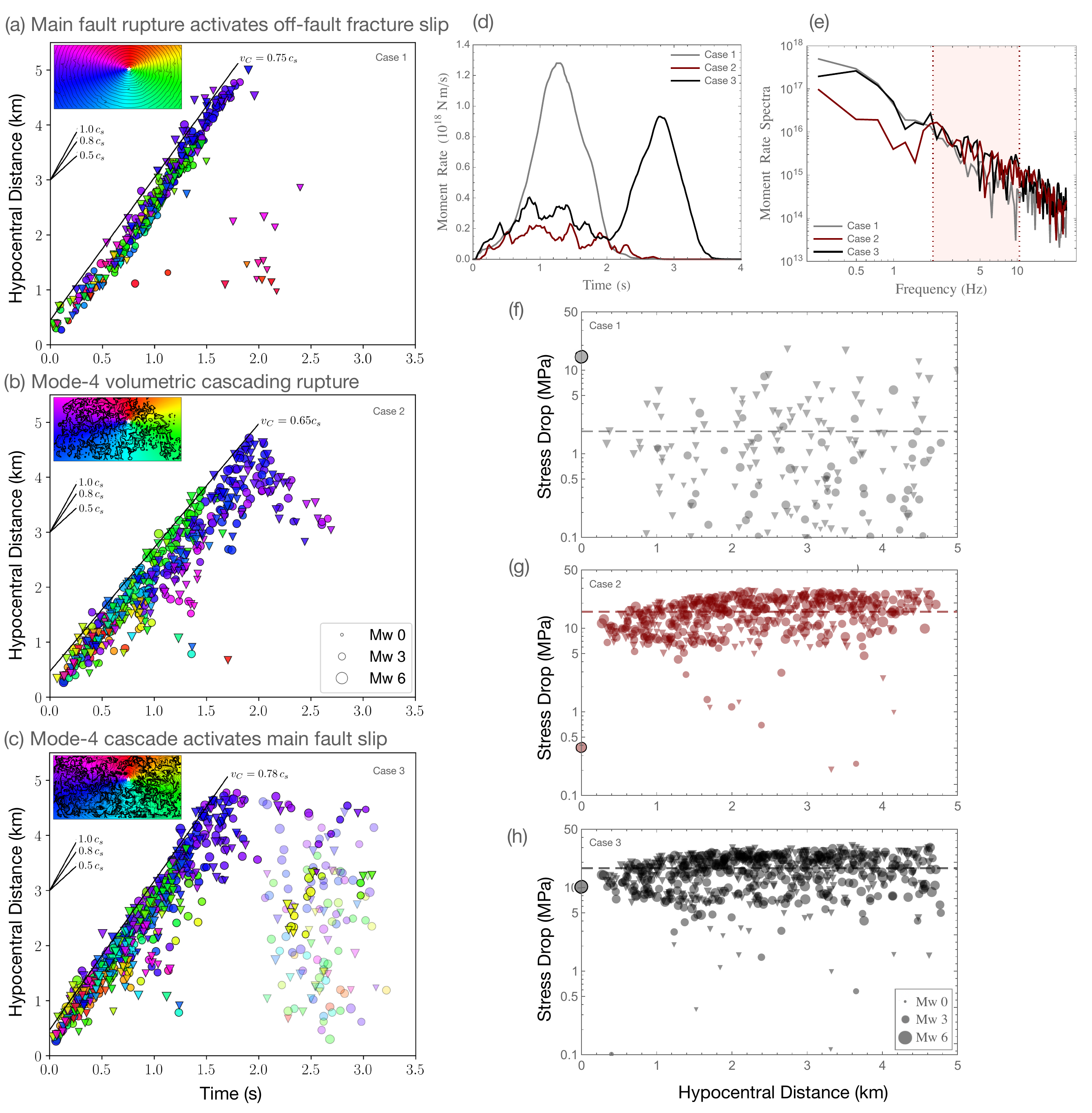}
\caption{
Kinematic and dynamic characteristics of multi-scale dynamic rupture earthquake cascades of Fig. \ref{fig:4}. Cascading speed ($v_C$, black line) for (a) Case 1: dynamic main fault rupture activating off-fault fracture slip, (b) Case 2: Mode-4 volumetric cascading rupture, and (c) Case 3: Mode-4 cascading rupture with delayed triggering of main fault slip. $v_C$ is the slipping fracture distance to the hypocenter versus its rupture onset time.
Triangles and dots represent fracture families 1 and 2, respectively, colors illustrate the azimuth of each slipping fracture with respect to the hypocenter. The size of each marker depicts the ruptured fracture's moment magnitude $M_w$. The top-left insets display rupture-time contours of the main fault rupture (contour interval: 0.08~s). The white star is the hypocenter. The background color indicates the azimuthal position relative to the hypocenter. The cascade rise time is $\sim 0.2$ s at a given hypocentral distance and azimuth (b-c), which is much shorter ($\sim 10\%$) than the overall duration of the cascading rupture. The transparent markers in (c) indicate off-fault slip during the later-stage main fault rupture on the fractures that have slipped previously as a part of the earlier cascade.
(d) Moment rate functions (MRF's) for Cases 1, 2, and 3 in grey, dark red, and black lines, respectively. 
(e) Moment rate spectra of the MRF's shown in (d). The highlighted frequency band $\approx$ 2-10 Hz corresponds to the estimated corner frequency range of the fracture population with radii from $R$ 100 to 500~m.
(f), (g) and (h) are the stress drop for each slipping fracture for Cases 1, 2, and 3, respectively. Triangles and dots represent fracture families 1 and 2, respectively. The dot at the hypocentral distance 0~km is the main fault value. The dashed line shows the spatial average of the fracture stress drop. 
}
  \label{fig:S3}
\end{figure*}

\begin{figure*}
\centering
\includegraphics[width=0.5\textwidth]{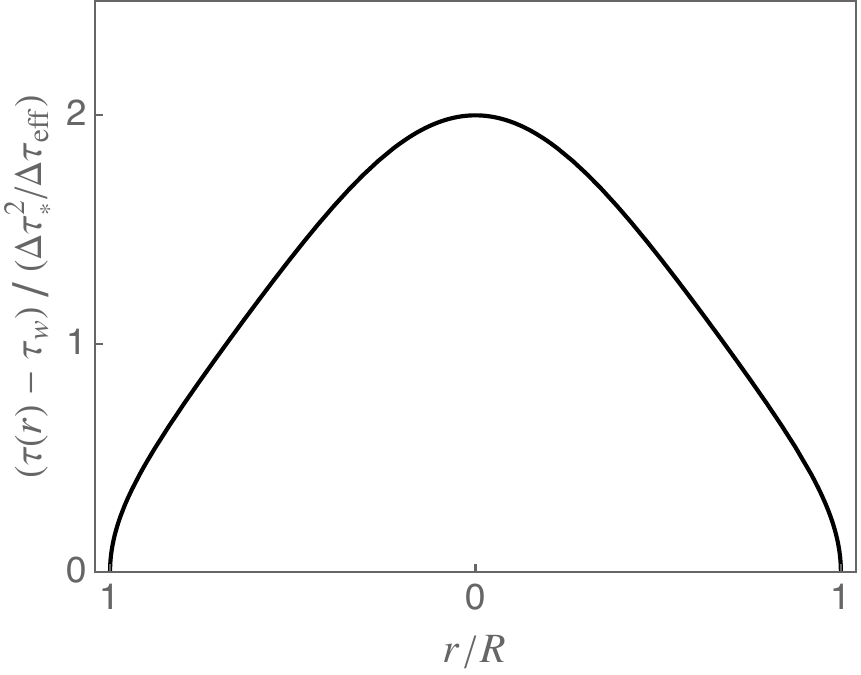}
\caption{Coseismic restrengthening $\tau(r,R)-\tau_{w}$ normalized by $\Delta\tau_{*}^{2}/\Delta\tau_{\text{eff}}$
as a function of the normalized position $r/R$ in our new analytical model of a crack-like, circular dynamic rupture on a fault with rate-state, flash-heating friction to estimate $G_{u/o}$. The peak restrengthening is reached at the hypocenter.}
  \label{fig:S4}
\end{figure*}

\begin{figure*}
\centering
\includegraphics[width=0.5\textwidth]{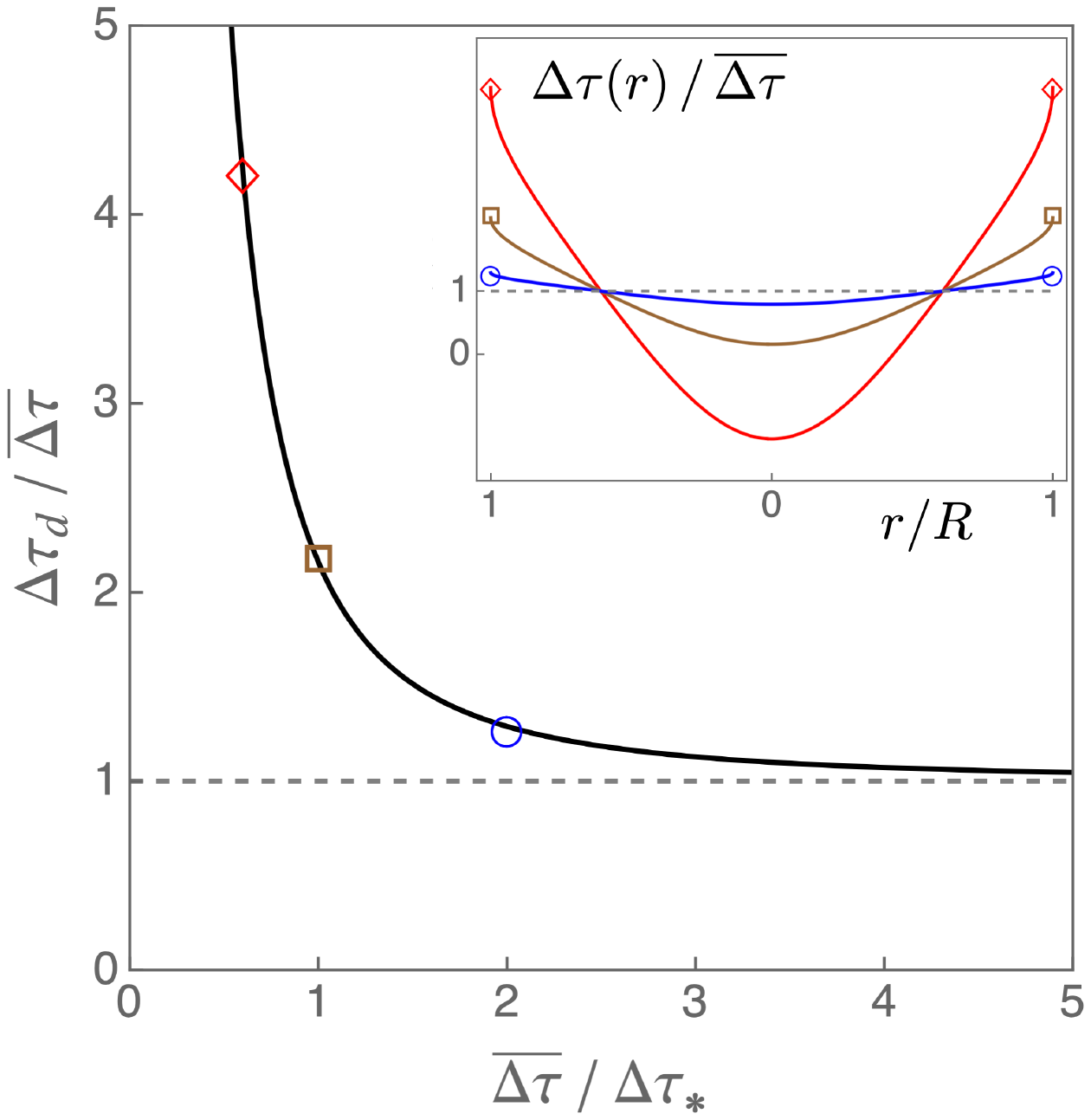}
\caption{Towards evaluating the fault-averaged $G_{u/o}$ from the analytical model. Maximum dynamic stress drop normalized by the rupture-averaged stress drop,  $\Delta\tau_d/\overline{\Delta\tau}$, as a function of the rupture-averaged stress drop normalized by the re-strengthening stress scale,  $\overline{\Delta\tau}/\Delta\tau_*$. The inset illustrates the model predictions of spatially-variable dynamic restrengthening, showing the distribution of the normalized stress drop, $\Delta\tau(r)/\overline{\Delta\tau}$, with the normalized radial distance $r/R$. The normalized stress drop varies from the maximum value ($\Delta\tau_d/\overline{\Delta\tau}$) near the front ($r/R=1$) to the minimum at the hypocenter ($r/R=0$).}
  \label{fig:SD}
\end{figure*}

\clearpage
\subsection*{Supplementary Tables}

\begin{table}[h]
    \centering
    \caption{Rate-and-state friction parameters used in the 3D dynamic rupture simulations.}
    \begin{tabular}{l c r}
    \hline 
    \textbf{Parameter} & \textbf{Symbol} & \textbf{Value} \\
    \hline
    Direct effect parameter     &       $a$     &   0.01 \\
    Evolution effect parameter  &       $b$     &   0.014 \\
    Reference slip velocity     &       $V_0$   &   $10^{-6}$ m/s \\
    Steady-state friction coefficient at $V_0$ & $f_0$ & 0.6 \\
    State evolution slip distance$^*$, $L$    &       $2\times10^{-5}\times R$     &   0.002 - 0.08 m \\
    Weakening slip velocity     &       $V_w$   &   0.1 m/s \\
    Fully weakened friction coefficient & $f_w$ & 0.1 \\
    Initial slip rate           &       $V_{ini}$ & $10^{-16}$ m/s \\
    \hline
    \multicolumn{3}{p{8.5cm}}{\footnotesize $^*$ For cases 1-3 in Sec. \ref{sec:5}. For case 4 in Materials and Methods, $L$ is constant $0.08$ m.}
    \end{tabular}
    \label{tab:friction-parameters}
\end{table}

\begin{table}[h]
\centering
\caption{Prestress and fault strength of fault and fractures for the 3D dynamic rupture simulation cases 1--4.}
\begin{tabular}{c c c c c c c}
\hline
\textbf{Case} & \boldmath{$S_{Hmax}$ ($^\circ$)} & \boldmath{$\mathcal{R}_0$} & \boldmath{$\mathcal{R}$ for the fault} & \boldmath{$\tau_0/\sigma'_n$ for the fault}
& \boldmath{$\mathcal{R}$ for fractures} & \boldmath{$\tau_0/\sigma'_n$ for fractures} \\
\hline
1 & 120 & 0.80 & 0.80 & 0.50 & 0.25 - 0.75 & 0.20 - 0.48 \\
2 & 65 & 0.80 & 0.77 & 0.48 & 0.25 - 0.80 & 0.20 - 0.50 \\
3 & 65 & 0.95 & 0.91 & 0.57 & 0.40 - 0.95 & 0.50 - 0.60 \\
4 & 65 & 0.80 & 0.77 & 0.48 & 0.25 - 0.80 & 0.20 - 0.50 \\
\hline
\end{tabular}
\label{tab:sim-parameters}
\end{table}

\clearpage
\section*{Materials and Methods}

\subsection*{Minimum nucleation size $R_c$ of a continuous coseismic weakening model under thermal pressurization}

We illustrate how previous explanations for the observed scaling of fracture energy fail to account for small-magnitude earthquakes. 
Continous coseismic slip weakening of fault strength driven by thermal pressurization (TP) at small slip is described by the undrained (adiabatic) limit $\tau=\tau_p\exp{(-\delta/\delta_c^{(TP)})}$ where $\tau_p$ is the peak strength and $\delta_c^{(TP)}=(\rho c/f\Lambda)h$ is Lachenbruch's slip weakening distance \cite{Lachenbruch80}.
To estimate the critical nucleation radius $R_c$, we adopt a linearized fault strength approximation, which is exact for ``critically-stressed'' faults \cite{GaragashGermanovich12} that reach dynamic instability at vanishingly small slip. \cite{UenishiRice2003,uenishi2018three} derive $R_c$ for a linear-slip-weakening circular crack as:
\begin{equation} \label{eq:RcTP}
R_c=1.299(\mu/\tau_p)\delta_c^{(TP)} \,.    
\end{equation}

We follow the parametrization of the TP model suggested by \cite{ViescaGaragash15}, which matches the seismologically inferred fracture energy versus coseismic slip data (Fig. \ref{fig:3}b, red line), with $\delta_c^{(TP)}=0.1$~m and $\tau_p=110$~MPa. Using these values, we infer a critical nucleation size of $R_c\approx 35$~m. Assuming a plausible stress drop of $\sim 10$~MPa, this TP model yields a minimum earthquake magnitude estimate of $M\approx2$.
This analysis highlights the limitations of the continuous coseismic weakening model under thermal pressurization in explaining the scaling of fracture energy for small earthquakes.

\subsection*{Scaling of state evolution slip distance and nucleation size with fault size}

We adopt the rate-and-state dependent model of high-velocity, flash heating weakening fault friction \cite{Noda09}, \cite{palgunadi2022dynamic}, \cite{dunham2011planar} which is employed throughout this study for energy balance considerations (Sec. \ref{sec:3}) and numerical modeling of cascading dynamic rupture (Sec. \ref{sec:5}). The model is parameterized as described in Table \ref{tab:friction-parameters}.

The rate-and-state fault strength breakdown near the rupture front can be effectively approximated by an exponential slip-weakening relation $\tau=\tau_{ss}+\Delta \tau_p\exp{(-\delta/L)}$, where $\tau_{ss}$ is the coseismic steady-state strength, $\Delta\tau_p=\tau_p-\tau_{ss}$ is the peak departure of strength from the steady-state, and $L$ is the state evolution slip distance \cite{Garagash2021,AmpueroRubin08}.
 The peak coseismic strength $\tau_p=f_p\sigma'_n$, where $f_p$ is the peak friction and $\sigma'_n$ is the normal effective stress, is only weakly dependent on the rupture front speed (Eq. 2.20, \cite{Garagash2021}) and can be approximated by a constant reference value $\tau_p \approx f_0\sigma'_n$ (Table \ref{tab:friction-parameters}). The coseismic steady-state friction, however, is expected to reach its fully weakened value, $\tau_{ss}\approx\tau_d\approx f_w\sigma'_n$, where $f_w$ is the corresponding fully weakened friction coefficient.

The corresponding fracture energy integral is given by $G=\int_0^{\delta_c}(\tau(s)-\tau_d)ds=(f_p-f_w)\sigma'_n L$, where the critical slip distance $\delta_c$, defined as the slip at which the strength is minimum (Fig. \ref{fig:2}a), is strictly infinite for exponential weakening. 
However, the energy breakdown is nearly complete, that is, $\tau\approx\tau_d$ for slip $> L$. 
Using parametrization $f_p=0.6$, $f_w=0.1$, and $\sigma'_n=40$ MPa (Table \ref{tab:friction-parameters}), we obtain for $\Delta\tau_p\approx (f_p-f_w)\sigma'_n\approx 20$ MPa and for the fracture energy $G\approx 20 \text{ [MPa]}\times L$.

Comparing this expression for the fracture energy with the seismologically-inferred scaling of the small-slip fracture energy, $G_c\approx 400\text{ [Pa]}\times R$ (Fig. \ref{fig:3}a), we derive the scaling of the state evolution slip distance as:
\begin{equation}\label{Methods:eq:L_R}
L=2\times10^{-5}\times R
\end{equation}

Considering the critical rupture nucleation size $R_c$ in the model with exponential slip weakening (Eq.\ref{eq:RcTP}), we find:
$R_c=1.299(\mu/\Delta\tau_p)L\approx 0.04\times R$.
This linear scaling of the small-slip fracture energy with ruptured fault size, established in this study, supports seismogenesis on all scales, that is, $R_c<R$ for all ruptured fault sizes $R$.

\subsection*{3D dynamic rupture simulations of earthquakes distributed across a damage zone fracture network and a main fault governed by scale-dependent fracture energy}

We use 3D dynamic rupture simulations and scale-dependent fracture energy to exemplify the dynamic interactions of coseismic, volumetric slip across a major fault interacting with a fracture network in its damage zone. 

\subsubsection*{Numerical model}

We assume a planar strike-slip fault, 8~km in length and 4~km in width, buried at a depth of 1~km. The fault strikes east-west with a dip of $90^\circ$ and is surrounded by a damage zone fracture network of 721 multi-scale fractures within a 10~km $\times$ 1~km $\times$ 6~km rock volume (Fig. \ref{fig:S1}a). We consider a homogeneous elastic isotropic material with P-wave speed $v_p=6$ km/s, S-wave speed $c_s=3.464$ km/s, rigidity $\mu = 32$ MPa, and bulk density $\rho = 2670$ km/m$^3$. 

The fracture network geometry is parameterized based on field observations and numerical models \cite{okubo2019dynamics,palgunadi2022dynamic},\cite{gabriel2021unified,dalguer2003simulation,ando2007effects} and generated using the commercial software FracMan. 
The resulting fracture network (Fig. \ref{fig:S1}) consists of two families with approximately conjugate average strikes of $25^\circ$ (382 fractures of family 1) and $-65^\circ$ (339 fractures of family 2) relative to the main fault and an average dip of $90^\circ$ (Fig. \ref{fig:S1}b). Fracture strike and dip each are characterized by a Fisher distribution \cite{fisher1995statistical} with a standard deviation of $\pm 10^\circ$. 

In nature, the maximum fracture length is limited by the width of the fault damage zone (e.g., \cite{MitchellFaulkner2009}).
In our model, the fracture density decreases with distance $r$ from the main fault following $\propto r^{-0.8}$ \cite{SavageBrodsky2011, palgunadi2022dynamic}, \cite{wang2005stereological}, and the size distribution of fractures, with radius $R$ between 100~m and 500~m, decays as a power-law $\propto R^{-2}$ (Fig. \ref{fig:S1}c, d) \cite{palgunadi2022dynamic},\cite{panza2018discrete,lavoine2019density}).
The fractures are assumed to be elliptical disks with an aspect ratio of 2 \cite{palgunadi2022dynamic},\cite{piggott1997fractal, berkowitz1998stereological}. About 75\% of all fractures are connected to more than one other fracture, 19\% are connected to only one fracture, and 6\% remain unconnected.

We use the open-source software SeisSol (\url{https://github.com/SeisSol/SeisSol}) to solve the 3D non-linearly coupled spontaneous dynamic rupture and seismic wave propagation problem with high-order accuracy in space and time. 
The computational domain spans a volume of 40~km $\times$ 40~km $\times$ 20~km, with an inner high-resolution cube centered at the main fault and its damage zone of 10~km $\times$ 1~km $\times$ 6~km. 
Within the high-resolution volume, the maximum element edge length is 250~m.
The resulting computational mesh consists of more than 40 million elements.
We use high-order basis functions of polynomial degree $p=4$, achieving $5^{th}$-order space-time accuracy for seismic wave propagation in all simulations. 

 Since the crack-tip process zone varies in size across fractures, we ensure to resolve the measured minimum process zone following  \cite{wollherr2018off} with two $5^{th}$-order accurate elements, equivalent to 12 Gaussian integration points. As a result, our smallest on-fault elements have an edge length as small as 4~m for the smallest fractures of size 100~m and their minimum dynamic rupture process zone size of 8~m. 
Such high-resolution 3D dynamic rupture simulations are computationally demanding. They are performed on the supercomputer Shaheen II (Cray XC-40), with a single high-resolution dynamic rupture simulation taking approximately 10 hours on 512 nodes.

The frictional strength $f\sigma'_n$ of the fault and fractures is governed by the same rate-state dependent model of high-velocity, flash-heating weakening \cite{Noda09}, \cite{dunham2011planar}. 
In our simulations shown in Sec. \ref{sec:5}, all friction parameters are considered constant across the fault and fracture network  (Table \ref{tab:friction-parameters}), except for the state evolution slip distance $L$, which scales with the fault and fracture size $R$, following our model of the small-slip fracture energy scaling with ruptured fault size (Eq. \ref{Methods:eq:L_R}). In Case 4 (Fig. \ref{fig:S2}), $L$ is set constant at 0.08~m for all fractures, the same as the main fault value.

In dynamic rupture models, only the minimum nucleation size $R_c$  needs to reach failure, while the remaining fracture or fault can be prestressed well below critical and yet break spontaneously. 
While dynamic rupture on most fractures in our models is initiated spontaneously by rupture branching or ``jumping'', the initial rupture nucleation involves prescribing a time-dependent over-stress \cite{harris2018suite} within a spherical volume with a radius of 400~m centered at a point on the fault (hypocenter). This over-stressed ``nucleation'' volume (indicated by the red dashed circle in Fig. \ref{fig:S1}a) intersects a small portion of the main fault and a subset of fractures.

The fault and fractures are loaded with the same ambient Cartesian stress tensor \cite{palgunadi2022dynamic} resembling an Andersonian \cite{Anderson1905} strike-slip background stress regime for all fractures and the main fault. The effective normal stress $\sigma'_n$ is depth-dependent and assuming hydrostatic conditions.
We define the stress ratio as $\phi = (\sigma_2 - \sigma_3)/(\sigma_1 - \sigma_3) = 0.5$ for our pure strike-slip regime, where $\sigma_1=S_{Hmax}$ is the maximum horizontal stress, $\sigma_2=S_v$ is the overburden stress, and $\sigma_3=S_{Hmin}$ is the minimum horizontal stress.
 We characterize the relative fault and fracture prestress by the relative prestress ratio $\mathcal{R}$, defined as the ratio of the maximum possible stress drop and frictional strength drop, $\mathcal{R} = \Delta \tau_d / (\tau_p - \tau_d)$ \cite{Taufiqurrahman2023}, \cite{Ulrich2019,aochi20031999}. 
$\mathcal{R}_0$ is the relative prestress ratio for a hypothetical,
optimally oriented fault location in the ambient stress field and thus defines an upper bound of $\mathcal{R}$ for our fault-fracture-network system under constant prestress. For $\mathcal{R} = \mathcal{R}_0$, the fracture or fault location is optimally oriented with respect to the local stress conditions, and for $\mathcal{R}_0 = 1$ an optimally oriented fault location is also critically stressed (e.g., \cite{Biemiller2022}).

We consider four dynamic rupture simulations across the same fault-fracture-network system, only varying the orientation of the ambient prestress $S_{Hmax}$ and the maximum relative prestress ratio ($\mathcal{R}_0$) (see Table \ref{tab:sim-parameters}). 
Cases 1, 2, and 4 feature overall sub-critical prestress with $\mathcal{R}_0 = 0.8$, while Case 3 allows for closer to critical prestress conditions with $\mathcal{R}_0=0.95$. 
The ambient stress orientation $S_{Hmax} = 120^\circ$ in Case 1 favors rupture on the main fault, which is characterized by local conditions of $\mathcal{R}= 0.8$ and $\tau_0/\sigma'_n = 0.5$. In distinction, $\approx90\%$ of all fractures are unfavorably prestressed with $\mathcal{R} \leq 0.3$ and $\tau_0/\sigma'_n \le 0.25$. In Cases 2, 3, and 4, where $S_{Hmax} = 65^\circ$ (Table \ref{tab:sim-parameters}), the ambient prestress favors both the main fault and fractures. 
Case 3 has a higher relative prestress ratio ($0.40 \le \mathcal{R} \le 0.95$) compared to Cases 2 and 4  ($0.25 \le \mathcal{R} \le 0.8$) reflecting its increased value of $\mathcal{R}_0$.

\subsubsection*{Kinematic and dynamic characteristics of multi-scale dynamic rupture earthquake cascades}

In Fig. \ref{fig:S3}, we examine the kinematic and dynamic properties of the three cases presented in Section \ref{sec:5}, with a focus on identifying observable signatures of cascading multi-scale ruptures in real fault zones.
We introduce the concept of a ``cascading rupture speed'' ($v_C$), defined as an apparent rupture speed across the 3D fracture network. $v_C$ is measured (Fig. \ref{fig:S3}a-c) as the distance to the hypocenter versus rupture onset time for each slipped fracture. In all three cases, we observe fractures from both families, located at various azimuths from the hypocenter, slipping (illustrated by colored symbols).

In Case 1, where the sustained main fault dynamic rupture activates off-fault slip (Fig. \ref{fig:4}a, Movies S1, S4), we find that $v_C = 0.75c_s$, which is approximately equal to the average rupture speed along the main fault (0.76$c_s$). 
This result reflects that fractures are dynamically activated by the passing main fault rupture front (inset in Fig. \ref{fig:4}a) in close proximity to the main fault. 
However, in Case 2 (Fig. \ref{fig:4}b, Movies S2, S4), the ``Mode-4'' volumetric cascading rupture speed is significantly lower at $v_C =0.65c_s$, reflecting a slower rupture cascade propagation due to rupture branching and dynamic triggering (``rupture jumping'') within the fracture network.

Case 2 exhibits the formation of a back-propagating cascade front towards the hypocenter after the rupture cascade has reached the outer edge of the fault zone at $t = 2$~s (Fig. \ref{fig:S3}b, Movies S2, S5). The volumetric dynamic cascade induces non-sustained main fault slip at several fracture-main-fault intersections, shown as densely spaced contours in the inset of Fig. \ref{fig:S3}b. In contrast, Case 3 has a cascading speed of $v_C = 0.78c_s$, comparable to Case 1 and the main fault rupture speed. In this simulation, the Mode-4 rupture cascade dynamically triggers sustained rupture on the main fault upon reaching the outer edge of the fault zone at $t = 2.1$~s. The ensuing main fault rupture renucleates slip on fractures directly connected to the main fault (indicated by transparent dots and triangles in Fig. \ref{fig:S3}).

In all three cases, we observe fracture-local supershear rupture speeds on smaller fractures immediately after rupture branches or jumps onto them from a larger fracture or the main fault, similar to findings in a different fracture-network main fault configuration \cite{palgunadi2022dynamic}.

In Fig. \ref{fig:S3}d-e, we analyze the dynamic rupture moment rate functions (MRFs) and their amplitude spectra. For Case 1, the MRF takes a simple triangle shape (grey line in Fig. \ref{fig:S3}d) akin to MRFs inferred for large natural earthquakes \cite{meier2017hidden}. In contrast, Case 2 leads to a more complex MRF characterized by multiple peaks (dark red line in Fig. \ref{fig:S3}d), each corresponding to slip on a large fracture within the cascade. The first part of Case 3 dynamic rupture, governed by volumetric cascading, also shows a multi-peaked moment rate release, but at a higher overall magnitude compared to Case 2, reflecting the higher prestress. Once the cascade activates sustained slip on the main fault rupture at $t = 2.1$~s, the MRF displays a triangle-shaped moment rate function similar to the main fault rupture in Case 1. 

Analyzing the moment rate spectra (Fig. \ref{fig:S3}e), we observe that cases with self-sustained rupture on the main fault (Cases 1 and 3) result in higher seismic moment release at lower frequencies. On the other hand, pure cascading Mode-4 rupture (Case 2) exhibits a depletion of low-frequency energy.
Those cases with cascading volumetric rupture episodes show larger amplitude high-frequency spectra. In Case 3, in particular, the frequency spectrum shows high energy release across all frequency ranges (0.01 - 13~Hz) due to its compound nature.
Notably, the frequency band $\approx$ 2 to 10~Hz, corresponding to the estimated corner frequency range of our fracture population with radii $R$ ranging from 100 to 500~m, shows less steep fall-off rates due to the ``Mode-4'' cascades in cases 2 and 3. 

Lastly, we explore the dynamic rupture examples' average stress drops in Fig. \ref{fig:S3}f-h, which align with observations of crustal earthquakes (e.g., \cite{huang2017stress}).  For Case 1, the average fracture stress drop is low (1.9~MPa), reflecting their relatively low prestress level (Table \ref{tab:sim-parameters}). Cases 2 and 3 have higher average fracture stress drop of 15.1MPa and 15.4~MPa, respectively. The average main fault stress drop in Cases 1 and 3 is approximately 10 MPa. Cases 1 and 3 fracture and main fault stress drops are an order of magnitude higher than the fracture stress drop for a pure ``Mode-4'' cascade (Case 2).
Our results compare well to the characteristics presented for a different fracture network realization, different main fault geometry, and different prestress,  and the detailed analysis in \cite{palgunadi2022dynamic}.

\subsection*{Analytical model of circular dynamic rupture with flash-heating friction}

In the context of the rupture energy balance (Fig. \ref{fig:2}a and Eq. \ref{Eq:G}), the fracture energy ($G$) can be decomposed into two components: the inferrable part ($G'$) and the non-inferrable part ($G_{u/o}$) from seismological observations in the form of seismic waveform spectra (e.g., \cite{Abercrombie2021}). 
Often, the non-inferrable part is neglected, leading to the assumption that $G \approx G'$ (e.g., \cite{AbercrombieRice05}).

To assess the validity of this common assumption, we examine Eq. \ref{Eq:G}) and observe that $G'$ is bounded from above by $\frac{1}{2}\Delta\tau\delta$, while $G_{u/o}$ is bounded by the dynamic under- or overshoot term $(\Delta\tau_d-\Delta\tau)\delta$. These bounds account for neglecting the negative contributions of the seismic radiation term $\mathcal{E}s$ to $G'$ and the coseismic restrengthening work $W_r$ to $G_{u/o}$
As a result, the condition for a negligible contribution of $G_{u/o}$ to $G$ can be expressed as $\Delta\tau_d-\Delta\tau \ll \frac{1}{2}\Delta\tau$, or alternatively, that dynamic stress under- or overshoot is significantly smaller than the final stress drop.

It is important to recognize that both crack-like and pulse-like ruptures driven by rate-dependent friction \cite{Noda09} \cite{Gabriel12} or thermal pressurization (e.g., \cite{Noda09} \cite{ZhengRice98, Garagash12, BrantutGaragashNoda19}) can lead to a coseismic rebound of stress, contributing to dynamic stress undershoot. This behavior potentially challenges the condition for $G_{u/o}$ to be negligible, thereby affecting the accuracy of estimating $G$.
In the following, we develop a novel analytical circular fault model with flash-heating friction to simulate coseismic restrengthening. This allows us to estimate a plausible coseismic-to-stress-drop relation and, consequently, determine $G_{u/o}$. By incorporating this information, we can refine the estimation of $G$ from seismological spectra for small events, offering a more accurate assessment of fracture energy.

\subsubsection*{Circular Rupture}

We consider a crack-like, circular dynamic rupture
on a fault with rate-state, flash-heating friction \cite{Noda09,Rice06}.

Two distinct regions characterize the rupture:

\emph{i. a near-front process zone:} This region near the rupture front experiences strength degradation from its peak value $\tau_{p}$, precipitated by the direct friction effect ahead of the advancing front. The strength then transitions to the flash-heating weakened steady-state value $\tau_{w}=\tau_{ss}(V\gg V_{w})$ over slip comparable to the state evolution slip distance $L$.

\emph{ii. the ``bulk'' of the rupture:} Away from the rupture front, the strength remains approximately at steady-state $\tau_{ss}(V)$ and recovers as the slip rate $V(r,t)$ decreases with distance from the rupture front. 

Such a structure of the solution requires the separation of scales between the process
zone and the rupture size \cite{Rice80,Barras20,garagash21}, which is achieved for
ruptures with large enough final slip $\gg L$.
 Under this condition, the near-tip process zone can be replaced by a crack tip singularity defined by the fracture energy of the breakdown process in the near-front process zone, $G_{c}\approx (\tau_{p}-\tau_{w})L$. The ``bulk'' of the dynamic rupture, from the hypocenter to the front, can be modeled as frictional steady-state behavior.

Therefore, for our analysis, we employ the self-similar singular Kostrov's solution \cite{kostrov1964selfsimilar} (also explored in \cite{dahlen1974ratio}) as an approximation for coseismic singular crack-like rupture propagation. We introduce a radial slip distribution ($0\le r\le R$) given by:
\begin{equation}
\delta(r,R)=\mathcal{C}(v_{r})\frac{\Delta\tau_{\text{eff}}}{\mu}\sqrt{R^{2}-r^{2}}\label{delta}\,,
\end{equation}
where $\Delta\tau_{\text{eff}}$ is the \emph{spatially-uniform} coseismic
stress drop,  $R$ is here the instantaneous rupture radius, and $v_{r}=dR/dt$
is the rupture speed. 
The coefficient $\mathcal{C}(v_{r})\sim1$ is given
by \cite{dahlen1974ratio} in the integral form (see their Eq.
(44) for a coefficient equivalent to our $(v_{r}/c_{s})\,\mathcal{C}(v_{r})$) and has the previously defined static limit $\mathcal{C}_{o}=\mathcal{C}(0)=(8/\pi)(1-\nu)/(2-\nu)$.

Here the original self-similar Kostrov's solution (corresponding to
constant $v_{r}$ and constant spatially-uniform $\Delta\tau_{\text{eff}}$)
is extended to approximate a rupture with slowly time-varying $v_{r}$
and $\Delta\tau_{\text{eff}}$.
We consider $\Delta\tau_{\text{eff}}$ as a measure of the actual, \emph{spatially-variable} stress drop \cite{garagash21} along the rupture driven by the flash-heating friction.

Away from the rupture front, we can then approximate the actual coseismic stress drop by assuming frictional flash-heating steady-state sliding, as $\tau=\tau_{ss}(V)$. 
Thus, we have:
\begin{equation}
\tau(r,R)\approx\tau_{ss}(V(r,R))=\tau_{w}+(\tau_{LV}-\tau_{w})\frac{V_{w}}{V(r,R)}\label{delta_tau}\,,
\end{equation}
where $\tau_{w}=f_{w}\sigma'_n$ and $\tau_{LV}=f_{LV}(V)\sigma'_n$ are expressed in terms of the fully weakened, minimum friction $f_w$ and the low-velocity steady-state friction $f_{LV}(V)$ \cite{Rice06, Noda09}.
For simplicity, and in view of the weak slip-rate dependence of $\tau_{LV}$, it
is further taken as a constant evaluated at a given representative coseismic
slip rate value. 

To complete our extended Kostrov model describing flash-heating driven dynamic rupture using the above approximate solution, we need to (i) provide the fracture propagation criterion to constrain $R(t)$ (referred to as the ``equation of motion'' \cite{Freund98,svetlizky2017EofM,weng2019dynamics,garagash21}), and (ii) evaluate the effective measure $\Delta\tau_{\text{eff}}$ of the stress drop distribution $\Delta\tau(r)$, as given by Eq. \ref{delta_tau}.

\subsubsection*{Propagation condition}

The propagation condition is a fundamental constraint on rupture dynamics, which states that the energy release rate averaged along the front of the circular rupture with instantaneous radius $R$, denoted as $\mathrm{ERR}_{\text{eff}}(R)$, must match the fracture energy $G_{c}(v_{r})$:
\begin{equation}
\text{ERR}_{\text{eff}}(R)=G_{c}(v_{r})\label{prop}\,.
\end{equation}

As provided by \cite{Madariaga76}, the total energy release required to propagate the Kostrov's rupture to a finite size $R$ is given by 
\[
E_{f}=\frac{\pi R^{3}}{3}\frac{\Delta\tau_{\text{eff}}^{2}}{\mu}g(v_{r})\,,
\]
where the function $g(v_{r})$ incorporates the dependence on the rupture speed \cite{Madariaga76},\cite{Ida72}.

By relating the total energy to the front-averaged energy release rate ($E_{f}=2\pi\int_{0}^{R}\text{ERR}_{\text{eff}}(r)rdr$) and considering that $\text{ERR}_{\text{eff}}(r)\propto r$, we can solve for the latter:
\begin{equation}
\text{ERR}_{\text{eff}}(R)=g(v_{r})\frac{\Delta\tau_{\text{eff}}^{2}R}{2\mu}\label{ERR}\,.
\end{equation}
In the limit of quasi-static ERR, the prefactor value $g(0)$ corresponds to $\mathcal{C}_{o}$.

\subsubsection*{Slip rate and stress distribution}

The slip rate distribution can be obtained by differentiating the slip (\ref{delta}) with respect to time: 
\begin{equation}
V(r,R)=\mathcal{C}(v_{r})\frac{\Delta\tau_{\text{eff}}\,v_{r}}{\mu}\mathcal{V}(r/R)\quad\text{with}\quad\mathcal{V}(\rho)=\frac{1}{\sqrt{1-\rho^{2}}}-\frac{1}{2}\sqrt{1-\rho^{2}}\label{V}\,.
\end{equation}
The above expression is derived by substituting the relation $d(\Delta\tau_{\text{eff}})/dt=-(v_{r}/2R)\Delta\tau_{\text{eff}}$, which follows from differentiating the propagation condition (Eq. \ref{prop}) with (Eq. \ref{ERR}) in time while assuming a slowly varying rupture speed, i.e., neglecting the time-derivatives of $v_{r}$ and $G_{c}(v_{r})$).

We use Eq. \ref{V} within Eq. \ref{delta_tau} to write for the coseismic stress distribution:
\begin{equation}
\tau(r,R)=\tau_{w}+\frac{\Delta\tau_{*}^{2}}{\Delta\tau_{\text{eff}}}\frac{1}{\mathcal{V}(r/R)}\label{Delta_tau}\,,
\end{equation}
where $\Delta\tau_{*}$ is a dynamic stress quantity defined as:
\begin{equation}
\Delta\tau_{*}^{2}=\frac{\mu\,(\tau_{LV}-\tau_{w})}{\mathcal{C}(v_{r})}\frac{V_{w}}{v_{r}}\label{Delta_tau_*}\,.
\end{equation}

Fig. \ref{fig:S4} illustrates the spatial distribution of the normalized coseismic restrengthening, given by $(\tau(r,R)-\tau_{w})/(\Delta\tau_{}^{2}/\Delta\tau_{\text{eff}})=1/\mathcal{V}(r/R)$ (Eq. \ref{Delta_tau}), from the fully-flash-heating weakened value $\tau_{w}$ attained at the rupture front. 
The maximum coseismic restrengthening occurs at the hypocenter ($r=0$), where $\tau(0)-\tau_{w}=2\,\Delta\tau_{}^{2}/\Delta\tau_{\text{eff}}$. Notably, the restrengthening effect $(\tau(r,R)-\tau_{w})/\Delta\tau_{}$ is inversely related to the effective (rupture-average) stress drop $\Delta\tau_{\text{eff}}/\Delta\tau_{}$, making it significant when the latter is not large.

\subsubsection*{Closing relation for effective stress drop $\Delta\tau_{\text{eff}}$ }

We can approximate the uniform coseismic stress drop $\Delta\tau_{\text{eff}}$, equivalent to the actual spatial-varying stress drop $\Delta\tau(r)=\tau_{o}-\tau(r)$ (Eq. \ref{Delta_tau}), using either of the following approaches:

(i) Matching the corresponding energy release rates of a static crack \cite{galis2015initiation,garagash21},

(ii) or, matching the mechanical work by equating $\Delta\tau_{\text{eff}}$ to the slip-weighted average of $\Delta\tau(r)$, i.e.:
\[
\Delta\tau_{\text{eff}}=\overline{\Delta\tau}:=\left\langle \Delta\tau\delta\right\rangle /\left\langle \delta\right\rangle \,.
\]
where $\left\langle ...\right\rangle =2R^{-2}\int_{0}^{R}(...)rdr$ denotes the fault-area average under axisymmetry.

For our purposes, we select the second approach as it is linked to considerations of the total energy released by the rupture \cite{noda2013stressdrop}.
The first approach is linked to the rate of energy release by propagating
rupture and seems appropriate when modeling rupture propagation \cite{garagash21,galis2015initiation,GalisAmpuero17}. 
Following (ii), we obtain 
\begin{equation}
\Delta\tau_{\text{eff}}=\overline{\Delta\tau}=\tau_{0}-\tau_{w}-B\frac{\Delta\tau_{*}^{2}}{\overline{\Delta\tau}}\,,\quad\text{with}\quad B=\frac{\int_{0}^{1}\frac{1}{\mathcal{V}(\rho)}\sqrt{1-\rho^{2}}\rho d\rho}{\int_{0}^{1}\sqrt{1-\rho^{2}}\rho d\rho}=3(\ln4-1)\approx1.159\label{dtau_ave}\,.
\end{equation}

\subsubsection*{Evaluating the fault-averaged $G_{u/o}$}

To evaluate $G_{u/o}$ averaged over the rupture of final radius $R$, we utilize the stress distribution derived earlier. 
We assume nearly complete flash-heating weakening near the rupture front ($\tau_{d}\approx\tau_{w}$) or, equivalently, that the peak dynamic stress drop is given by:
\begin{equation}
\Delta\tau_{d}\approx\tau_{0}-\tau_{w},\label{dtau_d}
\end{equation}
which we can then relate to the average coseismic stress drop $\overline{\Delta\tau}$
using Eq. \ref{dtau_ave}. 
This relation in the form of $\Delta\tau_{d}/\overline{\Delta\tau}-1=1.159\times(\Delta\tau_{*}/\overline{\Delta\tau})^{2}$ is shown in Fig. \ref{fig:SD}. 
The spatial distribution of the normalized stress drop $\Delta\tau(r)/\overline{\Delta\tau}$,
which follows from Eqs. \ref{Delta_tau} and \ref{dtau_d} and the above
expression relating $\Delta\tau_{d}$ and $\overline{\Delta\tau}$,
is shown in the inset for three values of the average stress drop (indicated by symbols).

Neglecting Madariaga's under- or overshoot, the final fault stress $\tau_{1}$ corresponds to the final coseismic stress $\tau(r)$ (Eq. \ref{Delta_tau}).
By substituting $\tau(r)$ and $\delta(r)$ with the approximate expressions for the circular fault given above in the integral for the first term in $G_{u/o}$, we obtain:
\begin{equation}
\left\langle (\Delta\tau_{d}-\Delta\tau)\delta\right\rangle =\frac{2}{R^{2}}\int_{0}^{R}(\tau(r,R)-\tau_{w})\delta(r,R)rdr=0.7726\times\mathcal{C}\frac{\Delta\tau_{*}^{2}}{\mu}R\label{under}
\end{equation}
with the exact prefactor being $2(\ln4-1)$. 

The restrengthening work at a point $r$ within the final rupture of radius $R$ is the integral of the stress excess over the fully dynamically weakened value of slip at that point.
Expressing the evolution slip at a point $r$ using the evolving rupture radius $R'$ from the moment when rupture first passes the point ($R'=r$) to the
final run-out $(R'=R)$, we can write for the restrengthening work
\[
W_{r}(r,R)=\int_{r}^{R}(\tau(r,R')-\tau_{w})\frac{\partial\delta(r,R')}{\partial R'}dR'=\int_{r}^{R}(\tau(r,R')-\tau_{w})\frac{V(r,R')}{v_{r}}dR'=\mathcal{C}\frac{\Delta\tau_{*}^{2}}{\mu}(R-r)\,.
\]
The fault-average restrengthening work is then given by:
\begin{equation}
\left\langle W_{r}\right\rangle =\frac{1}{3}\mathcal{C}\frac{\Delta\tau_{*}^{2}}{\mu}R\label{Wr}\,.
\end{equation}

In summary, we find that:
\begin{equation}
G_{u/o}=0.4393\times\tau_{*}R\quad\text{with}\quad\tau_{*}=\mathcal{C}\frac{\Delta\tau_{*}^{2}}{\mu}=(\tau_{LV}-\tau_{w})\frac{V_{w}}{v_{r}}\label{Methods:EQ:Guo}\,.
\end{equation}
Remarkably, $G_{u/o}$ such expressed in terms of the source size $R$ is independent of the average coseismic stress drop $\overline{\Delta\tau}$ and therefore of the details of the rupture propagation and the fracture energy of the flash heating process expanded near the rupture front. 
This independence arises because the flash-heating weakened strength excess over the minimum (fully flash-heated) strength, $\tau-\tau_{w}$, is inversely related to $\overline{\Delta\tau}$.
Therefore, the ``work''-intensity $(\tau(r)-\tau_{w})\delta(r)$ is independent of $\overline{\Delta\tau}$.

An alternative, equivalent expression for $G_{u/o}$ in terms of the average stress drop $\overline{\Delta\tau}$ and average slip $\left\langle \delta\right\rangle =\frac{2}{3}\mathcal{C}\frac{\overline{\Delta\tau}}{\mu}R$, can be obtained as follows:
\begin{equation}
G_{u/o}=0.6589\times\frac{\Delta\tau_{*}^{2}}{\overline{\Delta\tau}}\left\langle \delta\right\rangle \label{Methods:EQ:Guo1}\,.
\end{equation}

The significance of $G_{u/o}=G-G'$ in the earthquake energy budget can be evaluated by comparing it to the energetic quantity $\overline{\Delta\tau}\left\langle \delta\right\rangle $.
The relation for the normalized $G_{u/o}/\overline{\Delta\tau}\left\langle \delta\right\rangle =0.6589\times(\Delta\tau_{*}/\overline{\Delta\tau})^{2}$ (
Fig. \ref{fig:2}b), suggests that $G_{u/o}$ is negligible, and
therefore, $G\approx G'$, for large stress drop events, $\overline{\Delta\tau}\gg\Delta\tau_{*}$.
Conversely, the fracture energy is underestimated by $G'$ for events
with stress drop comparable or smaller than $\Delta\tau_{*}$.

For the parameters used in this study, $f_{LV}\approx0.545$ which is the low-velocity steady-state strength evaluated at $V\sim1$ m/s, $f_{w}=0.1$,
$V_{w}=0.1$ m/s, $\bar{\sigma}=51$ MPa, and $v_{r}=0.8c_{s}$ with
$c_{s}=3464$ m/s, and $\mu=32$ GPa, we find:
\[
\Delta\tau_{*}\approx5.0\text{ MPa},\quad\tau_{*}\approx820\text{ Pa},\quad G_{u/o}\approx 360\text{[Pa]}\times R.
\]

Thus, $G_{u/o}$ becomes an important contributor to the rupture energy balance for events with a stress drop not larger than $\approx5$ MPa.
This physics-based estimate of $G_{u/o}$ is then used to estimate the full fracture energy $G=G'+G_{u/o}$ from the seismologically inferred values of $G'$ for the events in the compilation of \cite{ViescaGaragash15} as shown in Fig. \ref{fig:3}. Not surprisingly, the lower bound of the data in Fig. \ref{fig:3}a emerges as the linear scaling with the source (ruptured fault) size $R$, which is consistent with the $G_{u/o}$ scaling established here.

Finally, it is worth noting that we here neglect Madariaga's overshoot, the potential difference between the final coseismic and the static stress due to the arresting phase of slip. For large stress drops, i.e.$>5$ MPa, Madariaga's overshoot would be non-negligible.
 \cite{Madariaga76}, \cite{Kaneko2014} estimates the size of the static stress drop to exceed the coseismic one by $\sim 20$\%.
 A corresponding correction for $G_{u/o}$ can be readily achieved by substituting the average coseismic value $\overline{\Delta\tau}$ in Eq. \ref{Methods:EQ:Guo1} with its expression in terms of the static value using above Madariaga's correction. Such a corrected Eq. \ref{Methods:EQ:Guo1} expressed in terms of the static stress drop would then have a  $\sim 20$\% larger numerical prefactor.

\bibliographyM{mode_4_refs}
\end{document}